\documentclass[lettersize, 11pt]{article}

\RequirePackage{amsthm,amsmath,amsfonts,amssymb}
\RequirePackage[authoryear]{natbib}
\usepackage[english]{babel}
\usepackage[utf8x]{inputenc}
\usepackage[T1]{fontenc}
\usepackage{diagbox}
\usepackage{amsmath}
\usepackage{amsfonts}
\usepackage{float}
\usepackage{graphicx}
\usepackage{bm}
\usepackage{mathtools}
\DeclarePairedDelimiter{\ceil}{\lceil}{\rceil}
\usepackage[colorinlistoftodos]{todonotes}
\usepackage[colorlinks=true, allcolors=blue]{hyperref}
\usepackage{color}
\usepackage[linesnumbered,ruled]{algorithm2e}
\usepackage[noend]{algpseudocode}
\AtBeginDocument{
  
}
\usepackage{cases}
\usepackage{multirow}
\usepackage{enumerate}
\usepackage{subfig}
\usepackage[export]{adjustbox}
\usepackage{afterpage}
\usepackage{lscape}

\newcommand{\bZ}{\bm{Z}}
\newcommand{\bY}{\bm{Y}}
\newcommand{\btheta}{\bm{\theta}}

\usepackage[a4paper,top=3cm,bottom=2cm,left=3cm,right=3cm,marginparwidth=1.75cm]{geometry}

\title{\bf Bayesian Bi-clustering Methods with Applications in Computational Biology}

\author{Han Yan$^1$\footnote{These authors contributed equally to the manuscript.} \and Jiexing Wu$^2$\footnotemark[\value{footnote}] \and Yang Li$^3$\footnotemark[\value{footnote}] \footnote{This presentation reflects the analysis and views of Yang Li. No recipient should interpret this presentation to represent the general views of Citadel Securities or its personnel. Facts, analysis, and views presented in this presentation have not been reviewed by, and may not reflect information known to, other Citadel Securities professionals.} \and Jun S. Liu$^1$}
\date{
    $^1$Department of Statistics, Harvard University\\
    $^2$Google\\
    $^3$Citadel Securities\\
}

\begin{document}
\maketitle
\begin{abstract}.
Bi-clustering is a useful approach in analyzing biological data when observations come from heterogeneous groups and have a large number of features. We outline a general Bayesian approach in tackling bi-clustering problems in moderate to high dimensions, and propose three Bayesian bi-clustering models on categorical data, which increase in complexities in their modeling of the distributions of features across bi-clusters. Our proposed methods apply to a wide range of scenarios: from situations where data are cluster-distinguishable only among a small subset of features but masked by a large amount of noise, to situations where different groups of data are identified by different sets of features or data exhibit hierarchical structures. Through simulation studies, we show that our methods outperform existing (bi-)clustering methods in both identifying clusters and recovering feature distributional patterns across bi-clusters. We apply our methods to two genetic datasets, though the area of application of our methods is even broader. 
\end{abstract}

\section{Introduction}
Over the past few decades, the biology community has witnessed an explosion of data due to technology advances. For example, DNA microarrays and the next-generation RNA sequencing technologies enable scientists to measure  expression levels of tens of thousands of genes at once; and the single-cell technologies can even extend such measurements to the individual cell level. 
Another example is the Human Genome Project, which collected the whole genome sequence information, i.e., the information  over three billion nucleotide base pairs that make up human DNA, from many thousands of individuals. Data collected in these cases are of high dimension in nature. Analyzing such data can reveal important biological functions, identify mutations that are responsible for certain types of diseases, shed light on human evolutionary  history, etc. A useful first attempt is to cluster data into homogeneous groups. Traditional clustering methods, such as the hierarchical clustering-based UPGMA \citep{upgma} and the distance-based K-means method \citep{kmeans, hartigan1979algorithm} have been successfully applied to biological data to identify groups of genes or samples that are of biological significance. However, the underlying biological mechanisms and the high dimensional nature of the data both suggest a further examination of the behavior of features while performing clustering. For example, when analyzing  gene expression data across cell types in the same tissue/organ, only a small fraction of the genes may be highly expressed and show different expression levels in different cell types. To cluster cells by cell types, it is advised to find out the differentially expressed genes and cluster cells only based on the behavior of those genes. Additionally, different biological functions or conditions activate different sets of genes. A gene may be highly expressed for a subset of experiments, while inactive in the rest of conditions. It is of great interest to identify samples corresponding to different biological conditions or functions and, at the same time,  identify the set of genes that are active under each condition or function. This motivates the need of bi-clustering, i.e., methods that  cluster both samples and features simultaneously. 

\subsection{A case study} \label{sec:case}
In an effort to detect population structures encoded by human single nucleotide polymorphisms (SNPs), we downloaded the human SNP data from the international HapMap project website, which made available the SNP information for thousands of individuals across the world. A SNP is a single nucleotide base-pair variation among populations, and is the most common type of sequence variations found in human DNA. The raw data consist of 1,397 individuals from 11 populations across the world, and 1,457,897 SNPs for each individual. Acknowledging that not all SNPs are population or continent specific 
we take a bi-clustering approach. 
Before formally introducing our methods and analyzing the real data, we demonstrate using a semi-synthetic dataset created from the human genetic data introduced above   potential advantages of bi-clustering over traditional clustering methods. 

We randomly select 50 individuals from each of the following four populations: two populations with Africa ancestry: Yoruba in Ibadan, Nigeria (YRI) and Maasai in Kinyawa, Kenya (MKK), one population in east Asia: Han Chinese in Beijing, China (CHB), and one population from Europe: Toscans in Italy (TSI). We keep 50 original SNPs (features) and add 450 noise features (SNPs generated randomly) with each generated from a probability vector following the Dirichlet distribution $Dir(1,1,1)$.
We apply to this dataset our biclustering method to be introduced in Section \ref{subsec: jiexing_model}, the K-means method based on the first three principal components (PCs) \citep{PCA}, and STRUCTURE \citep{pritchard2000inference}, a clustering method that treats each SNP as cluster-differentiating. Clustering results are shown in Table \ref{tab: motExp}. With only 50 informative SNPs and a signal to noise ratio of $\frac{1}{9}$, our bi-clustering method achieves the highest adjusted Rand Index (ARI, \citep{ari_hubert}) of 0.68, followed by 0.41 of STRUCTURE, and 0.40 of the PCA+Kmeans method. ARI is a measure of similarity between two data clustering results after correcting for chance. Its maximum value is 1 and occurs when two clustering results match perfectly. All three methods perform reasonably well in separating populations by continents. However our bi-clustering method is least affected by noise as fewer samples from each underlying population is misclassified into other clusters. All methods have difficulties in distinguish the two African populations, as they are geographically and genetically close to each other. PCA+Kmeans creates two clusters for the two populations, but with both populations almost equally mixed in both clusters. STRUCTURE puts almost all individuals from both African populations into one cluster. In contrast, our bi-clustering method can still put the individuals into two clusters with each cluster dominated by one underlying population.

\begin{table}[h!]
\begin{center}
\begin{tabular}{|l | c c c c| c c c c| c c c c|}
\hline
Methods&  \multicolumn{4}{c|}{Bi-clustering} & \multicolumn{4}{c|}{PCA+Kmeans} & \multicolumn{4}{c|}{STRUCTURE}\\
\hline
\diagbox{TrueLabel}{EstLabel} & C1 & C2 & C3 & C4 & C1 & C2 & C3 & C4 & C1 & C2 & C3 & C4\\
\hline
 CHB & 47  & 1& 2 & 0   & 36 &3& 4  &  7& 39 & 2& 9 & 0\\
 \hline
TSI  &  1 & 49 & 0 & 0  & 1  & 43&0 &  6  &  4& 40& 3& 3\\
\hline
 MKK &  1 & 1 &  34 &14 & 1  & 1 &27& 21 & 2&  4&42& 2\\
\hline
YRI  &  0 & 0& 11  & 39 & 2  & 0&27 & 21  & 1 &  1& 48& 0\\
\hline
\end{tabular}
\end{center}
\caption{Clustering results by three methods using the semi-synthetic dataset.}
\label{tab: motExp}
\end{table}

\subsection{Literature on bi-clustering methods}\label{intro}
The idea of bi-clustering was first introduced in \cite{hartigan1972direct}. \cite{cheng2000biclustering} pioneer the application of bi-clustering to study gene expression data. They define a bi-cluster as a submatrix of the expression data matrix (possibly transformed and normalized) that have a high similarity score, where their similarity score reflects how well the submatrix can be fitted by a two-way ANOVA model. They use a greedy algorithm to search for bi-clusters that have high scores. In \cite{bergmann2003iterative}, bi-clusters are termed as ``transcription modules'' that contain coregulated genes in a set of relevant experimental conditions. Mathematically, a transcription module contains genes whose normalized expression levels exceed a threshold under all conditions that belong to the bi-cluster. Statistical-Algorithmic Method for Bi-cluster Analysis (SAMBA) is a graph based approach proposed by \cite{tanay2002discovering}. They define a bipartite graph whose two stable sets correspond to genes and conditions, with edges for significant expression changes. Bi-clusters correspond to bipartite subgraphs. They assign weights to vertex pairs so that significant bi-clusters have large weights. \cite{ben2003discovering} work with the relative ordering of gene expression values, and define bi-clusters to be order-preserving sub-matrices (OPSMs). A submatrix is order preserving if there is a permutation of its columns under which the sequence of values in every row is strictly increasing.

Another set of approaches utilizes singular value decomposition (SVD) to discover bi-clusters. A rank-$k$ data matrix $\bm{Y}$ with dimensions $n$ and $p$ can be decomposed into the sum of $k$ rank-$1$ matrices: $\bm{Y}=\bm{U\Lambda V^T}=\sum\limits_{i=1}^k s_i\bm{u}_i\bm{v}_i^T$, where $\Lambda=diag(s_1,\cdots,s_k)$, $s_1\ge \cdots \ge s_k>0$ are $k$ singular values, $\bm{U}=(\bm{u}_1,\cdots,\bm{u}_k)$ are orthonormal left singular vectors and $\bm{V}=(\bm{v}_1,\cdots,\bm{v}_k)$ are orthonormal right singular vectors. \cite{RoBiC} use the rank-1 matrix $s_1\bm{u}_1\bm{v}_1^T$ to approximate $\bm{Y}$. Applying a hinge function, a bi-cluster is extracted by taking out rows and columns whose corresponding entries in $\bm{u}_1$ and $\bm{v}_1$ have large absolute values. Once a bi-cluster is found, the same procedure is applied to the remaining matrix $Y-s_1\bm{u}_1\bm{v}_1$. Instead of using a hinge function, \cite{lee2010biclustering} directly impose a sparsity constraint on the top singular vector and define a bi-cluster to be rows and columns that have non-zero entries in $\bm{u}_1$ and $\bm{v}_1$. 

A useful feature of SVD is that if two columns of $\bm{Y}$ are the same, the corresponding columns in $\bm{V}^T$ must also be equal. A similar observation applies to the rows of $\bm{Y}$. This property can be used to find bi-clusters that have constant values. \cite{hierarchicalbiclustering} combine this idea with hierarchical clustering to cluster left and right singular vectors. Rows and columns in the same cluster define a constant bi-cluster of the original data matrix. 

The plaid model introduced by \cite{lazzeroni2002plaid}  treats each expression value in a bi-cluster as a sum of the main effect, the gene effect and the condition effect. If a gene-condition pair is in multiple bi-clusters, its expression value is the sum of module effects from all the bi-clusters it falls into. Bi-clusters are found by minimizing mean squared error through an iterative algorithm. Inspired by the plaid model, two Bayesian bi-clustering algorithms were developed about the same time:  \cite {gu2008bayesian} and  \cite {caldas2008bayesian}. The main difference between the two methods is that  \cite{gu2008bayesian} assume that the expression value of a gene in a bi-cluster is centered around the sum of three terms: the bi-cluster-specific main effect, the gene effect, and the condition effect. Those genes that do not fall into any bi-cluster have their expression values following a normal distribution with mean 0. In contrast, in \cite{caldas2008bayesian}, the cluster-specific main effect term is replaced by a global main effect term that is shared by every observation regardless of whether the data point belongs to any cluster or not.

All aforementioned bi-clustering methods are designed for continuous observations. We also find a few methods that target or may be modified to target  categorical data. \cite{neuwald2003ran} introduce a Bayesian Partition with Pattern Selection model to simultaneously cluster aligned protein sequences in a super family into  sub-families and discover important positions in the aligned sequences that underlie these subfamilies.
\cite{patrikainen2004subspace}  formulate a bi-clustering task on binary data through a finite mixture model and term it ``subspace clustering''. Positing that the Bernoulli parameter for each feature equal either a feature-specific background value or a cluster-specific value if the feature is relevant for a bi-cluster $k$, they develop an EM-like algorithm to fit the model and use Chi-squared tests and Bayesian Information Criterion (BIC) \citep{schwarz1978estimating} to select features for each cluster and the total number of clusters. \cite{hoff2005subset} puts a Bayesian spin on the model of \cite{patrikainen2004subspace} by imposing a Dirichlet process mixture model. A version on continuous data is discussed in \cite{hoff2006model}.
\cite{SpecCo} propose the SpecCo algorithm  to find a pre-specified $K$ number of clusters to partition the input binary data matrix based on  a generalized modularity measure. 

An idea similar to bi-clustering is co-clustering or block clustering, which partitions both rows and columns of a dataset. For example, \cite{LBM14Keribin} partition both rows and columns into an unknown number of clusters. Each row-cluster and column-cluster pair form a data block. Within each block, elements are homogeneous, i.e., following the same distribution. One underlying assumption of block clustering is that all features are of the same nature, which greatly limits the applicability of 
block clustering. 

Some non-probabilistic criterion-based algorithms for  bi-clustering of categorical data have also been proposed.
\cite{pensa2005bi_boolean} analyze the Boolean context of a data matrix. SUBCAD in \cite{gan2004highD_Categorical} finds bi-clusters iteratively by optimizing a combination of ``compactness'' within a bi-cluster and ``separation'' outside bi-clusters. \cite{hash2016} uses Locality Sensitive Hashing (LSH) to find an initial set of seed clusters and then uses the  InClose \citep{Inclose} algorithm from Formal Concept Analysis to find bi-clusters.

\subsection{Main themes of the article}
We here use the term ``bi-clustering'' broadly and do not distinguish between subspace clustering and bi-clustering, thus including both  unsupervised learning (aka clustering) with variable selection and clustering with cluster-specific feature selection. For example, BBC2 introduced in Section \ref{subsec: jiexing_model} assumes for each feature  a feature-specific background distribution and a set of cluster-specific foreground distributions depending on whether the feature is selected by any cluster. Some Bayesian methods for  clustering with variable selection have been developed in the early 2000s \citep{liu2003bayesian,neuwald2003ran,tadesse2005bayesian,raftery2006variable}.They are mostly model-based and aim to use a small subset of features to partition objects into different clusters. Each selected feature is usually assumed to follow the same distribution within a cluster but different distributions for different clusters. In contrast, when we allow for cluster-specific feature selection, the goal is to identify subsets of objects and features simultaneously. This strategy may give us more freedom in modeling features and  greater interpretability along both dimensions.

In some studies, data from different sources collected under different conditions need to be integrated. In these cases, bi-clustering may be more appropriate given that certain biological functions are gene-specific and condition-specific. In some other studies, samples are more similar to each other: e.g., experiments conducted under the same condition and from the same tissue. The subset of genes that are active are likely to be the same across groups. Then, one may want to perform a variable selection to identify this subset. BBC1 to be introduced in Section \ref{subsec: basic} falls under clustering with variable selection. 

The model we develop in Section \ref{sec: general framework} shares some similarity with the multiple partitions model in \cite{multi-clsuter_modelbased} and \cite{multi-partitionClustering}. The multiple partitions mixture model assumes that features can be grouped into an unknown $B$ number of independent blocks.  Within each block, observations over the set of features follow a mixture model with an unknown number of mixture components. The emphasis of multi-partitions clustering is that no single partition of the objects can fully describe the behavior of the data over all features. Two samples can be in the same cluster over a set of features, but in two different clusters over another set of features. Our bi-clustering model 
can also be viewed as a special multi-partitions clustering model if in each feature block we view the partition at row-cluster level, but not individual sample level. Our model also restricts how row-clusters interact in each feature block. A mixture component either represents a single row-cluster or a group of row-clusters.

We  propose a general framework for bi-clustering problems in the Bayesian context in Section \ref{sec: general framework}. Although our framework works for various data types, we present  in Section~\ref{sec3} detailed treatments of a few special twists aiming at different tasks in biology. Specifically, three methods with increasing complexity are introduced to tackle categorical data. The first method (BBC1) works on binary data, and performs a Bayesian clustering with variable selections. We then  model a more complicated situation by allowing a feature to have cluster-specific distributions in only some of the clusters (BBC2). Compared with BBC1, BBC2 is able to identify both features that are unique to a cluster, and clusters that share certain features. We further extend BBC2 to build a hierarchy of clusters (HBBC), which is able to detect finer differences in the distributions of features among clusters compared with BBC1 and BBC2. For continuous data, we describe a  general framework for detecting bi-clusters based on independent correlation matrices of the features. This enables us to integrate datasets from different sources and identify consistent patterns among datasets. Applications of the proposed methods on a few genetic and genomic datasets are presented in Section \ref{application: HapMap}. We conclude with a short discussion in Section \ref{sec: conclusion}. More technical details are provided in the Appendix.

\section{A general framework for Bayesian bi-clustering}\label{sec: general framework}

\subsection{A statistical model}
Let $\bm{Y}$ denote the $n \times p$ data matrix. For consistency, we call rows of $\bm{Y}$ objects, and columns features. The goal is to assign objects into clusters, and for each cluster, find a subset of features that supports the cluster assignment, i.e., features that follow cluster-specific distributions. 
The parameters of interest are: the total number of clusters $K$, the latent cluster assignment vector 
$\bm{C}=(C_1,...,C_n)^T$, and the $K\times p$ binary feature selection matrix $\bm{S}=(S_{k,j})_{k=1,j=1}^{K,p}$. Note that $\bm{C} \in \{0,1,\cdots,K\}^n$, where $0$ represents a null group hosting objects that are not assigned to any cluster and $k>0$ represents cluster $k$. $S_{k,j}=1$ means that feature $j$ is included in cluster $k$. For each $k>0$, let $I_k=\{i: i\in\{1,\cdots,n\}, C_i=k\}$ and $J_k=\{j: j\in\{1,\cdots,p\},S_{k,j}=1\}$. The submatrix $\{Y_{i,j}\}_{i\in I_k,j\in J_k}$ form a bi-cluster. Conditioning on all the parameters, each element of $\bm{Y}$ follows a posited parametric model $F$, and we use $\bm{\theta}_{k,j}$ to denote model parameters for feature $j$ under cluster $k$:
\begin{align*}
Y_{i,j} \mid  \bm{\bm{C}, \bm{S}, \Theta}, K \sim 
F\left[\bm{\theta}_{c_i,j}\mathbb{I}\left(S_{c_i,j}=1\right)+\bm{\theta}_{0,j}\mathbb{I}\left(S_{c_i,j}=0\right)\right].
\end{align*}
where $\bm{\Theta}=\{\bm{\theta}_{0,j},\bm{\theta}_{1,j},\cdots,\bm{\theta}_{K,j}\}_{j=1}^p$.
This model assumes that the distribution of $Y_{i,j}$ depends on its cluster assignment $C_i$ and whether feature $j$ is selected as a distinguishing feature for cluster $C_i$. Parameter vectors $\bm{\theta}_{c_i,j}$ and  $\bm{\theta}_{0,j}$ represent the cluster-specific and feature-specific distributions of feature $j$, respectively. 
Conjugate priors for elements of $\bm{\Theta}$ are particularly convenient to work with in this setting since the posterior distributions are still from the same distribution family, and the normalizing constants can be calculated explicitly. The $C_i$'s  are often assumed to be  independently and identically distributed ($iid$) as:
\begin{equation}\label{prior I}
P(C_i=0 \mid K)=\gamma_0, \quad \text{and} \quad\; P(C_i=k \mid K)=\left(1 - \gamma_0\right) / K, \ \hspace{2mm} k=1,...,K, 
\end{equation}
where  $\gamma_0$ is the prior probability for an object to belong to the null cluster, and can be set to 0 if we wish for a partition of all objects. As a binary variable, $S_{k,j}$ is assumed to be independent Bernoulli {\it a priori}:
\begin{equation} \label{prior S}
S_{k,j} \mid  K \sim \text{Bernoulli}(\pi_S), \quad j=1,\dots,p,\ \  k=1,\dots,K.
\end{equation}
In (\ref{prior S}), all the $S_{k,j}$'s share the same prior distribution. This is reasonable if one does not have any prior belief that a particular subset of features are more likely to be differently distributed across clusters. If one wants to incorporate prior knowledge on feature importance, one can replace $\pi_{S}$ with feature specific parameters $\pi_{S_j}$, where $\pi_{S_j}$ is set larger for more importance features. 
The prior in (\ref{prior S}) gives rise to all feature distribution patterns across clusters. In the special case of clustering with variable selection, only two patterns are considered for each feature $j$: $S_{k,j}=0$ for all $k$ and $S_{k,j}=1$ for all $k$. In this case $\bm{S}$ degenerates to a $p$-dimensional binary vector.

Similar discrete-data bi-clustering problems and models have also been studied in \cite{neuwald2003ran}, \cite{Guo}, and \cite{Jiexing}, although in more specific ways. Our general framework here can be viewed as an extension of their frameworks.

\subsection{Posterior inference via Gibbs sampler}
Our parameters of interest are $\bm{C}$, $\bm{S}$, and $K$. The dimension of $\bm{S}$ depends on $K$. The prior distribution of $\bm{C}$ also depends on $K$. Theoretically we can use a reversible jump MCMC algorithm \citep{green1995reversible} for posterior computation, but it is computationally expensive with poor convergence behaviors. We therefore opt for a less principled but computationally more friendly two-step approach: for a range of $K$ values, we make inference on  $\bm{C}$ and $\bm{S}$ conditional on $K$; and then we choose the optimal $K$ according to a certain criterion, such as maximizing an approximation of  $P(K\mid \bm{Y})$. In the current subsection, we focus on the inference of $\bm{C}$ and $\bm{S}$, and will discuss how to find the optimal $K$ in the next subsection.

It is straightforward to write down the joint posterior distribution of $\bm{C}$ and $\bm{S}$ conditioning on $K$:
\begin{align} \label{Post_General}
\begin{split}
P(\bm{C},\bm{S}\mid \mathbf{Y},K) &=\frac{P(\bm{Y\mid C},\bm{S},K)P(\bm{C}\mid  K)P(\bm{S} \mid K)}{P(\mathbf{Y} \mid K)}\\
& \propto P(\bm{Y \mid C},\bm{S},K)P(\bm{C} \mid K)P(\bm{S} \mid K).
\end{split}
\end{align}
We  will mainly focus on finding the posterior modal estimates of the model parameters 
because of the multimodality of the posterior distribution resulting from near-nonidentifiabilty of typical clustering and bi-clustering models.
Due to high dimensionality and the complexity of  the parameters and latent variables involved in the model, it is infeasible to obtain either an analytical or a guaranteed numerical optimal solution. Instead, we can apply Markov Chain Monte Carlo (MCMC) methods to draw samples from the posterior distribution to identify an approximate mode of (\ref{Post_General}).

Conditioning on $K$, we run a Gibbs sampler to iteratively draw elements in $\bm{C}$ and $\bm{S}$ conditional on the rest. By the Bayes rule, the posterior probability of selecting feature $j$ for cluster $k$ is,
\begin{align}\label{post S}
\begin{split}
 & P\left(S_{k,j}=1\mid\boldsymbol{C},\boldsymbol{S}_{[-(k,j)]},\mathbf{Y},K\right)\\
= & \frac{\pi_{S}P\left(\mathbf{Y}\mid\boldsymbol{C},S_{k,j}=1,\boldsymbol{S}_{[-(k,j)]},K\right)}{\pi_{S}P\left(\mathbf{Y}\mid\boldsymbol{C},S_{k,j}=1,\boldsymbol{S}_{[-(k,j)]},K\right)+\left(1-\pi_{S}\right)P\left(\mathbf{Y}\mid\boldsymbol{C},S_{k,j}=0,\boldsymbol{S}_{[-(k,j)]},K\right)},
\end{split}
\end{align}
where $\bm{S}_{[-(k,j)]}$ denotes the feature selection matrix with the $(k,j)$-entry omitted. When $K$ is not too large, we can also calculate $P(\bm{S}_j \mid \bm{C},\bm{S}_{[-j]} , Y, K)$, where $\bm{S}_{[-j]}$ denotes the submatrix of $\bm{S}$ with the $j$-th column removed, for all configurations of $\bm{S}_j=(S_{1,j},\cdots,S_{K,j})^T$, and sample from it.
The conditional distribution of each element of $\bm{C}$ is categorical with probability:
\begin{align} \label{post I}
\begin{split}
P(C_i =k \mid \bm{C}_{[-i]},\bm{S},\bm{Y},K) \propto P(\mathbf{Y}\mid C_i=k,\bm{S},\bm{C}_{[-i]},K) P(C_i = k \mid K),
\end{split}
\end{align}
for $ k=0,\dots,K$, where $\bm{C}_{[-i]}=(C_1,C_2,\cdots,C_{i-1},C_{i+1},\cdots,C_n)^T$.

We obtain the maximum  {\it a posteriori} (MAP) estimate of $(\bm{C},\bm{S})$ from the MCMC samples. For any given $K$, running the Gibbs sampler $M$ times after burn-in gives us $(\bm{C}^1(K),\bm{S}^1(K)),...,(\bm{C}^M(K),\bm{S}^M(K))$. The MAP estimate is defined as
\begin{align}\label{eq: CS_MargY}
\begin{split}
& (\bm{\hat{C}}(K),\bm{\hat{S}}(K)) \\
&= \underset{ m=1,...,M} {\arg \max} \ P(\bm{C}^m(K),\bm{S}^m(K) \mid \mathbf{Y},K)\\
& =\underset{ m=1,...,M} {\arg \max} \ P(\mathbf{Y}\mid \bm{C}^m(K),\bm{S}^m(K),K)P(\bm{C}^m(K)\mid K)P(\bm{S}^m(K)\mid K).
\end{split}
\end{align}

We sometimes can further integrate out $\bm{S}$  to obtain the marginal posterior distribution of  $\bm{C}$ and $K$ up to a normalizing factor, if $K$ is not too large. This marginalization can further improve the efficiency of the Gibbs sampler as suggested in \cite{liu1994collapsed}. Since the features are  independent {\it a priori} and all entries in $\bm{S}$ are binary, we have 
\begin{align}\label{MAP_C_marginal}
\begin{split}
P\left(\mathbf{Y}\mid\bm{C},K\right) &=\sum_{\bm{S}}P\left(\mathbf{Y}\mid\bm{C},\bm{S},K\right)P\left(\bm{S}\mid K\right)\\
& =\prod_{j=1}^p \sum_{\bm{S}_j} P\left(\mathbf{Y}_{\cdot,j}\mid\bm{C},\bm{S}_j,K\right)P\left(\bm{S}_j\mid K\right),
\end{split}
\end{align}
where $\bY_{\cdot,j}=(Y_{1,j},Y_{2,j},\cdots,Y_{n,j})^T$.
This leads to the marginal  MAP estimate of $\bm{C}$, which can also be approximated using MCMC samples:
\begin{align}\label{eq: C_margY}
\bm{\tilde{C}}(K) = \underset{m=1,...,M} {\arg \max} \ P(\mathbf{Y}\mid \bm{C}^m(K),K)P(\bm{C}^m(K) \mid K).
\end{align}

After obtaining $\bm{\tilde{C}}(K)$, we can proceed
to get an estimate of $\bm{S}$:
\begin{align}\label{eq: Stilde}
\begin{split}
\tilde{\bm{S}}(K) = \underset{ m=1,...,M} {\arg \max} \ P(\mathbf{Y}\mid \bm{S}^m(K), \bm{\tilde{C}}(K),K)P(\bm{S}^m(K) \mid K),
\end{split}    
\end{align}
or get the ``exact''  posterior mode of $\bm{S}$ by evaluating $P(\mathbf{Y}\mid \bm{S}(K), \bm{\tilde{C}}(K),K)P(\bm{S}(K) \mid K)$ for all configurations of $\bm{S}(K)$.

In our bi-clustering model, we do not assume a mixture membership for each object.
Since cluster labels are of categorical nature without a natural ordering, calculating their posterior means or medians is not appropriate here. Hence we use the MAP estimate in (\ref{eq: CS_MargY}) or (\ref{eq: C_margY}) as a point estimate of the latent cluster membership. 
Although it may induce biases for estimating the model parameters if we fix the latent membership indicators at their MAP estimates,
the biases are typically not a major concern because
the posterior distribution is mostly highly concentrated at the MAP in such cases.
For example, for the human genetic data we analyzed in Section \ref{sec: bbc2 hapmap}, 1,195 out of 1,198 (99.75\%) latent cluster assignment variables are concentrated at their respective posterior mode throughout the sampling steps after the burn-in stage. For the single cell RNA sequence data we analyzed in Section \ref{subsec: scRNAseq}, each $C_i$ takes the value of its posterior mode in all MCMC steps after burn-in. These real data examples imply that the posterior probability for the latent cluster assignment variable $C_i$ to be at its posterior mode is very close to 1. 
Furthermore, the parameters of primary interest to us are $\bm{C}$ and $\bm{S}$ but not $\bm{\Theta}$, and we integrate out $\bm{\Theta}$ in our posterior computation and inference.

\subsection{Inference of number of clusters $K$}\label{sec:K}
To infer the number of clusters $K$, a natural approach is to choose  $K$ that maximizes the marginal posterior probability $P(K \mid  \mathbf{Y})$, i.e.,
\begin{align*}
\hat{K}=\underset{K}{\arg\max}\,P\left(\mathbf{Y}\mid K\right) P\left(K\right),
\end{align*}
which requires the marginalization computation:
\begin{align} \label{Marginal Likelihood}
\begin{split}
P(\mathbf{Y} \mid K)
=\sum_{\bm{C}} \sum_{\bm{S}} P(\mathbf{Y}\mid \bm{C,S},K) P(\bm{C}\mid K) P(\bm{S}\mid K).
\end{split}
\end{align}

Equation (\ref{Marginal Likelihood}) generally does not have an analytical solution. We can approximate it using \cite{chib1995marginal}'s method, which will be briefly reviewed below, with its specific application to our task deferred to Section \ref{subsec: jiexing_model}. 

Using $\bm{\theta}_K$ to denote the set of parameters in the model, whose dimension may be related to $K$, and letting $\bZ$ be the latent variable (or missing data) in the model, we can write $P(\mathbf{Y} \mid K)$ as
\begin{equation}\label{BMI}
P(\mathbf{Y} \mid K)=\frac{P(\mathbf{Y}\mid \bm{\theta}_K, K) P(\bm{\theta}_K\mid K)}{P(\bm{\theta}_K \mid \mathbf{Y}, K)}.
\end{equation}
where $P(\bm{\theta}_K \mid K)$ is the prior density of $\bm{\theta}_K$, and $P(\mathbf{Y}\mid \bm{\theta}_K, K)=\int P(\mathbf{Y},\bZ\mid \bm{\theta}_K, K) d\bZ$. Equation \eqref{BMI} holds for any $\bm{\theta}_K$.  However, to estimate $P(\bm{\theta}_K \mid \mathbf{Y}, K)$ in the denominator accurately, it is advised to choose a $\bm{\theta}_K$ with high posterior density, such as the posterior mean or mode computed from the Gibbs outputs. We denote the selected $\bm{\theta}_K$ value as  $\bm{\theta^*}_K$.

Suppose Gibbs sampling can be applied to sample $\btheta_K$ and $\bZ$ from their conditional densities: $P(\bm{\theta}_K \mid \bm{Y},\bm{Z}, K)$ and $P(\bm{Z}\mid \bm{Y}, \bm{\theta}_K,K)$. Let $\{ \bm{\theta}_K^{(m)},\bm{Z}^{(m)} \}_{m=1}^M$ be $M$ such samples. We obtain the approximation
\begin{equation}\label{chibs1}
 \hat{P}(\bm{\theta}^*_K \mid \mathbf{Y}, K) =\frac{1}{M}\sum_{m=1}^{M}P\left(\bm{\theta^*_K}\mid\bm{Z}^{\left(m\right)},\mathbf{Y}, K\right).
\end{equation}
Note that the likelihood function in the numerator of Equation (\ref{BMI}) can be written as 
\[ P(\bY\mid\btheta_K,K) =\frac{P(\bY,\bZ\mid \btheta_K,K)}{P(\bZ\mid \bY,\btheta_K,K)}.\]
In many examples, such as when the missing data distribution $P(\bZ \mid \bY, \btheta_K,K)$ and the complete-data likelihood $P(\bY,\bZ\mid \btheta_K,K)$ are analytically available, the numerator of Equation (\ref{BMI}) can be evaluated exactly and, thus,  
 (\ref{BMI}) can be approximated by using $\hat{P}(\bm{\theta}^*_K \mid \bY, K)$ in the denominator.
When $P(\bZ \mid \bY, \btheta_K,K)$ is not available analytically, 
we will need to  find some way to approximate it \citep{chib1995marginal}.

In practice, computing marginal likelihoods for all possible values that $K$ can take, i.e., from 1 to $N$, and choose the best among them is computationally demanding and not realistic. Narrowing down the potential values to a smaller range is the first step. To achieve this, we can first run HBBC, a model proposed in Section \ref{subsec: hbbc} that can automatically infer the number of clusters, denoted as $\hat{K}_{HBBC}$. With $\hat{K}_{HBBC}$, we can run other bi-clustering algorithms proposed in Section \ref{sec3} for a sequence of consecutive $K$ values around $\hat{K}_{HBBC}$, and select the best model using the marginal likelihood as the criterion. Without resorting to HBBC, assuming we know {\it a priori} $K\in[K_{min},K_{max}$], we can first compute the approximate posterior $P(K \mid \bY)$ for a small candidate set of ordered $K$ values (possibly equally spaced) in between $K_{min}$ and $K_{max}$. Identify the value $K=K_1$ whose approximate $P(K \mid \bY)$ is the largest. Let $K_2$ and $K_3$ be the two values in the candidate set that are immediately before and after $K_1$. Replace $K_{min}$ by $K_2$ and $K_{max}$ by $K_3$. Repeat the same procedure till we narrow down the search to a small, manageable set of candidate values. 
In many examples we encounter, $P(K\mid Y)$ increases with $K$ first, reaching its maximum and then decreases. Hence the above approach ensures that the optimal $K$ is always inside the narrowed searching interval. If there are concerns that $P(K\mid \bm{Y})$ is not unimodal, We may also need to consider the neighborhood of those $K$s whose $P(K\mid \bm{Y})$s are not significantly smaller than the maximum $P(K\mid \bm{Y})$. In practice, we may take $K_{min}=1$ and $K_{max}=[n/20]$, for example.

\section{Bayesian bi-clustering with various data types}\label{sec3}
{Besides the massive amount of gene expression data, there is also an abundance of discrete data available in genetic and genomic studies. For example, the Human Genome Project has made available complete or partial DNA sequences and genetic variations of many individuals of different population origins, making it possible to study genetic bases of various diseases and the human migration history. Protein sequences and structural data for multiple species available from the Protein Data Bank make it possible to analyze similarities and differences among  proteins in a superfamily, revealing insights on their functionalities.
In this section we introduce three Bayesian bi-clustering methods with increasing complexity that target categorical datasets, and a novel bi-clustering model for data integration.}

\subsection{Bayesian clustering with variable selection for binary data}\label{subsec: basic}
Binary data matrices arise in many applications and are also ideal prototypes for developing relevant theories. For example, in text analyses, columns correspond to words (or concepts) and rows correspond to articles. Bi-clustering can provide a complementary view to the classic analysis based on topic models \citep{blei2003latent}.  Such data can also be generated from single-cell RNA sequencing technologies, in which a ``1'' means that the gene is highly expressed in the cell, and ``0'' otherwise. Since experimented single cells are often from the same tissue, we usually expect only a small fraction of genes to show different expressed/repressed patterns among cell sub-types.  The clustering problem then becomes a bi-clustering one: identifying differently expressed genes and clustering cells based on these genes.

To make our presentation more targeted, we refer to rows of $\bm{Y}$  as ``cells'', columns  as ``genes'', and selected columns as ``biomarkers''. A simplifying assumption that we adopt here is that  biomarkers have cluster-specific distributions across all clusters and non-biomarkers assume only feature-specific distributions. Thus, the feature selection matrix $\bm{S}$ in the general framework is simplified to a $p$-dimensional vector, $\bm{S}=\left(S_1,\dots,S_p\right)^T \in \{0,1\}^p$, with $S_j$= 1 indicating that gene $j$ is selected as a biomarker, and 0 otherwise. Biomarkers are assumed to have different probabilities of being in the expressed state in each cluster, while each non-biomarker follows its feature-specific background distribution across all clusters. The model, termed as ``BBC1'', is thus:
\begin{align}\label{eq:basic}
\mbox{BBC1:} \ \  \ \ \  Y_{i,j} \mid  \bm{ C, S, \Theta}, K \sim 
\text{Bernoulli} \left(\bm{\theta}_{c_i,j}\mathbb{I}\left(S_{j}=1\right)+\bm{\theta}_{0,j}\mathbb{I}\left(S_{j}=0\right)\right).
\end{align}

The priors for $C_i$ and $S_j$ follow (\ref{prior I}) and (\ref{prior S}) and the priors for the $\theta$'s are i.i.d. $\text{\text{Beta}}\left(\alpha_{\theta_1},\alpha_{\theta_2}\right)$  for biomarkers, and i.i.d. $\text{\text{Beta}}\left(\alpha_{w_1},\alpha_{w_2}\right)$ for non-biomarkers.
The marginal likelihood for BBC1 after integrating out $\bm{\Theta}$ is,
\begin{align*}
P\left(\mathbf{Y}\mid\boldsymbol{C},\boldsymbol{S}, K \right) = \left[\prod_{j:\,S_{j}=0}\frac{\text{B}\left(\alpha_{w_1}+n_{j,1},\alpha_{w_2}+n_{j,0}\right)}{\text{B}\left(\alpha_{w_1},\alpha_{w_2}\right)}\right]\left[\prod_{j:\,S_{j}=1}\prod_{k=1}^{K}\frac{\text{B}\left(\alpha_{\theta_1}+n_{k,j,1},\alpha_{\theta_2}+n_{k,j,0}\right)}{\text{B}\left(\alpha_{\theta_1},\alpha_{\theta_2}\right)}\right],
\end{align*}
where $n_{j,0}=\sum_{i}\mathbb{I}\left\{ Y_{i,j}=0\right\}$, $n_{j,1}=\sum_{i}\mathbb{I}\left\{ Y_{i,j}=1\right\}$, $n_{k,j,0}=\sum_{i}\mathbb{I}\left\{ C_{i}=k,Y_{i,j}=0\right\}$, $n_{k,j,1}=\sum_{i}\mathbb{I}\left\{C_{i}=k,Y_{i,j}=1\right\}$  and $\text{B}\left(\cdot,\cdot\right)$ is the beta function. After incorporating the priors, we sample $\bm{C}$ and $\bm{S}$ from their joint posterior distribution via Gibbs sampling. To accelerate the convergence of the Gibbs sampler, we can further integrate out $\bm{S}$  if either clustering is the major concern, or we want to get the marginal MAP estimate of $\bm{C}$ first, and  infer $\bm{S}$ conditioning on it:
\begin{align}\label{eq: Basic_MargLike}
\begin{split}
& P\left(\mathbf{Y}\mid\boldsymbol{C}, K\right) \\
= & \sum_{S_{1}\in\left\{ 0,1\right\} }\cdots\sum_{S_{p}\in\left\{ 0,1\right\} }P\left(\mathbf{Y}\mid\boldsymbol{C},\boldsymbol{S}, K\right)P(\bm{S} \mid K)\\
= & \prod_{j=1}^{p}\left[\left(1-\pi_{S}\right)\frac{\text{B}\left(\alpha_{w_1}+n_{j,1},\alpha_{w_2}+n_{j,0}\right)}{\text{B}\left(\alpha_{w_1},\alpha_{w_2}\right)}+\pi_{S}\prod_{k=1}^{K}\frac{\text{B}\left(\alpha_{\theta_1}+n_{k,j,1},\alpha_{\theta_2}+n_{k,j,0}\right)}{\text{B}\left(\alpha_{\theta_1},\alpha_{\theta_2}\right)}\right].
 \end{split}
\end{align}

Combining  Equation (\ref{eq: Basic_MargLike}) with the prior in (\ref{prior I}), we have the posterior distribution for $\bm{C}$, from which we can  sample using Gibbs sampling. If a single optimal solution for $\boldsymbol{C}$ is preferred, as it is easier to interpret and proceed to further experimental investigations, the sample MAP estimator (i.e., the one obtained via MCMC samples) can be used. 
After inferring $\bm{C}$, we proceed to derive the posterior distribution of $\bm{S}$:
\begin{align*}
    S_j \mid \bm{C},\bm{Y}, K \sim \text{Bernoulli}\Bigg(\frac{\pi_S \prod\limits_{k=1}^K \frac{B(\alpha_{\theta_1}+n_{k,j,1},\alpha_{\theta_2}+n_{k,j,0})}{B(\alpha_{\theta_1},\alpha_{\theta_2})}}{\pi_S \prod\limits_{k=1}^K \frac{B(\alpha_{\theta_1}+n_{k,j,1},\alpha_{\theta_2}+n_{k,j,0})}{B(\alpha_{\theta_1},\alpha_{\theta_2})}+(1-\pi_S)\frac{B(\alpha_{w_1}+n_{j,1},\alpha_{w_2}+n_{j,0})}{B(\alpha_{w_1},\alpha_{w_2})}} \Bigg),
\end{align*}
for $j=1,\ldots,p$.

To infer the number of cell sub-types, $K$, we select the number that maximizes the marginal likelihood $P(\bY | K)$, which can be written as:
\begin{align}\label{margLike:basic}
    \begin{split}
        P(\bY \mid K)=\frac{P(\bY \mid \bm{C},K)P(\bm{C} \mid K)}{P(\bm{C} \mid \bY, K)}.
    \end{split}
\end{align}
The two terms in the numerator are readily available from (\ref{prior I}) and  (\ref{eq: Basic_MargLike}) for any realization of $\bm{C}$. The denominator can be approximated by the empirical frequency of  the posterior samples at $\bm{C}$ given $K$. To have a good approximation to the denominator, we evaluate the term at the sample posterior mode $\bm{C}^*$. 

For any $\bm{C}$, a permutation of the cluster indices $1,2,\cdots,K$ generates an equivalent bi-cluster pattern. Denoting the equivalent class as $\tilde{\bm{C}}$, we have $P(\bm{C} \mid \bY, K)=\frac{1}{K!}P(\tilde{\bm{C}} \mid \bY, K)$. Each equivalent pattern in $\bm{\tilde{C}}$ is a local mode for the Gibbs sampler. As a result, our sampling result may serve as a representation not for the posterior distribution of $\bm{C}$ but for that of $\bm{\tilde{C}}$. Let $(\bm{\tilde{C}}^m)_{m=1}^M$ be outputs from the Gibbs sampler, and $\tilde{\bm{C}}^*$ represents the sample posterior mode and its equivalent patterns. We approximate $P(\bm{\tilde{C}}\mid \mathbf{Y},K)$ at $\tilde{\bm{C}}^*$ by its frequency among the $M$ samples.

The model is not strictly identifiable as permuting the labels in $\bm{C}$ does not change the model likelihood. \cite{allman2009identifiability} proved that finite mixture of $K$ different Bernoulli products are ``generically identifiable'', i.e., the set of non-identifiable parameters is of Lebesgue measure zero, up to label switching if the number of Bernoulli components $p$ satisfies $p \ge 2\ceil{\log_{2}K}+1$.  Due to the presence of non-biomarkers, we cannot directly apply this result here. However, it is not difficult to show directly using Theorem 4 and Corollary 5 of \cite{allman2009identifiability} that if the number of biomarkers $p_1\equiv \sum_j S_j \ge 2\ceil{\log_{2}K}+1$, BBC1 is generically identifiable. Conditioning on $K$ and $\bm{S}$, model parameters for biomarkers are generically identifiable. For non-biomarkers, all clusters share the same set of model parameters, which are strictly identifiable. Combining the two together, BBC1 is generically identifiable, up to label switching. When both $\bm{S}$ and $\bm{\Theta}$ are unknown, the model is also generically identifiable. Two different $\bm{S}$ vectors give rise to two different distributions since the set of features that are independent of each other is different under different configurations of $\bm{S}$.




To test the method, we simulate datasets from model (\ref{eq:basic}) with $n=200$, $p=1,000$, $K=5$, 
and the number of biomarkers $N_s$=10, 20, 30, and 40. Each setting is replicated 20 times. In each dataset, we randomly pick $N_s$ biomarkers and assign each object to one of the $K$ clusters with equal probability. For a non-biomarker $j$, we  draw $\theta_{0,j} \sim \text{Beta}\left(1,1\right)$, and for a biomarker $j$, we draw $\theta_{k,j} \sim \text{Beta}\left(0.2,0.2\right)$, $k=1,\ldots,5$. 

We compare performances of BBC1, FABIA: factor analysis for bicluster acquisition \citep{fabia}, and two Latent Class Analysis with variable selection approaches in \cite{LCA2017_2}, termed as  ``VarSelLCM-BIC'' and ``VarSelLCM-MICL'' respectively, implemented in the R package \textit{VarSelLCM} \citep{LCA2017_1}.  FABIA is a popular biclustering method for continuous data. Adding in a comparison with FABIA, we demonstrate that biclustering methods that are designed for continuous data may not perform well on categorical data. Conditioning on  model parameters and the number of clusters, both approaches in \textit{VarSelLCM} assume a mixture model: biomarkers follow cluster-specific distributions and non-biomarkers follow feature-specific distributions in all clusters. ``VarSelLCM-BIC'' employees a non-Bayesian approach, using the EM algorithm \citep{EM1977} to obtain the MLE and selecting the best model based on the BIC. ``VarSelLCM-MICL'' is a Bayesian approach similar to BBC1. Model selection in this case is based on maximizing $P(\bY,\bm{C} \mid \bm{S},K)$.    

We implement BBC1 with the following parameter specifications. $\theta_{k,j}\sim Beta(1,1)$ for $k=0,1,\cdots,K$, and $j=1,\cdots,p$. The prior belief on the proportion of biomarkers is $\pi_S=0.1$. We initialize $\bm{C}$ and $\bm{S}$ at random, run the MCMC chain for 900 steps, and discard the first 200 steps as burn-in. We let $K=2,3,\cdots,9$ and choose the best $K$. The running time depends on the signal level. When the number of biomarkers is 10 or 20, it takes 4 to 6 minutes to finish analyzing one dataset. When the number of biomarkers is 30, the average running time is 30 seconds. The times further reduce to less than 15 seconds when the number of biomarkers is 40.

Results are shown in Table \ref{tab: sim_basic}. The clustering error (CE) rate refers to the misclassification error rate after optimal alignment between the inferred and true cluster assignments. BBC1 outperforms the other methods. The false positive rate and false negative rate refer to those for features. BBC1 can always find the actual number of clusters, with no or very low clustering errors. Most biomarkers and non-biomarkers can be correctly classified. VarSelLCM-BIC almost always estimates the number of clusters to be 2. In many cases, the two estimated clusters are mixed with objects from all 5 underlying groups. The high false negative rates indicate that it cannot find biomarkers. The performance for VarSelLCM-MICL increases significantly as we increase the number of biomarkers. The model for VarSelLCM-MICL is very similar to BBC1 except for some prior and hyper-parameter specifications. However, it employs an iterative {\it ad hoc} optimization strategy to estimate $\bm{C}$ and $\bm{S}$. 
Briefly, with a fixed $K$ it finds $\bm{S}$ to maximize
$P(\bm{S}\mid \bY,K ,\bm{C})$ conditional on $\bm{C}$, and then 
conditional on $\bm{S}$, it randomly selects some components of $\bm{C}$ and updates them by
maximizing $P(\bY,\bm{C} \mid \bm{S},K)$.
However, without a complete sweep, some elements of $\bm{C}$ may seldom get updated, resulting in clustering errors even when $N_s$ is large. When $N_s$ is too small, the algorithm is also easily stuck in a local mode.
FABIA performs the worst among all methods. Under the model setup for FABIA, all data entries that do not belong to any bi-cluster follow the same distribution. Hence we treat such data as being in a null cluster. The estimated number of clusters for FABIA is the number of bi-clusters identified plus a null cluster. Most of the time, it cannot find any bi-cluster. Its high false negative rates indicate that it misses out many signal features. Its high false positive rates implies that many noise features are wrongly picked up to infer bi-clusters.

\begin{table}[H]
\begin{center}
\begin{tabular}{|l |c |c c c c c |}
\hline
Number of & Method & Number of & CE & ARI & False Positive & False Negative\\
biomarkers & & Clusters & Rate & & Rate & Rate \\
\hline
\multirow{4}{*}{Ns = 10} & BBC1& 5 & 3.45\% & 0.93 & 0.46\% & 12\%\\
\cline{2-7}
& VarSelLCM-BIC & 2 & 72.53\% & 0.04 & 2.95\% & 78.5\%  \\
\cline{2-7}
& VarSelLCM-MICL & 4.3 & 59.4\% & 0.22 & 20\% & 38\% \\
\cline{2-7}
& FABIA & 1.7 & 75.93\% & 0 & 35.85\% & 80\% \\
\hline
\hline
\multirow{4}{*}{Ns = 20}& BBC1 & 5 & 0.025\% & 1 & 0.17\% & 2.5\% \\
\cline{2-7}
& VarSelLCM-BIC & 2 & 71.38\% & 0.06 & 2.44\% & 90.75\% \\ 
\cline{2-7}
& VarSelLCM-MICL & 3.6 & 31.48\% & 0.62 & 2.90\% & 9\% \\
\cline{2-7}
& FABIA & 1.4 & 76.48\% & 0 & 35.34\% & 74.75\% \\
\hline
\hline
\multirow{4}{*}{Ns = 30}& BBC1 & 5 & 0 & 1 & 0.14\% & 2.18\% \\
\cline{2-7}
& VarSelLCM-BIC & 2 & 65.33\% & 0.17 & 2.53\% & 95.5\% \\ 
\cline{2-7}
& VarSelLCM-MICL & 5.1 & 15.98\% & 0.83 & 1.47\% & 2.34\% \\
\cline{2-7}
& FABIA & 1.65 & 76.4\% & 0 & 35.69\% & 75.79\% \\
\hline
\hline
\multirow{4}{*}{Ns = 40}& BBC1 & 5 &  0 & 1 & 0.13\% & 2.88\%\\
\cline{2-7}
& VarSelLCM-BIC & 2 & 64.83\% & 0.18 & 2.26\% & 99.88\% \\ 
\cline{2-7}
& VarSelLCM-MICL & 5.05 & 11.78\% & 0.87 & 0.21\% & 0.89\%\\
\cline{2-7}
& FABIA & 1.6 & 74.95\% & 0.01 & 36.18\% & 73.35\% \\
\hline
\end{tabular}
\end{center}
\caption{Simulation results to testing out the effectiveness of Bayesian bi-clustering based on BBC1 for binary data.}
\label{tab: sim_basic}
\end{table}

\subsection{Bayesian bi-clustering on categorical data}\label{subsec: jiexing_model}
The model in the previous section
assumes that a feature follows either a  feature-specific background distribution (the same for all clusters), or a cluster-and-feature-specific distribution, which is different for different clusters. However, in some scenarios a feature may follow cluster-specific distributions only in some clusters, and follow the feature-specific background distribution in the remaining clusters. To deal with such scenarios, we introduce  BBC2, which is largely based on the framework detailed in \cite{Jiexing}.

Let $\bm{Y}$ be an $n\times p$ matrix with entry $Y_{i,j}$ taking value from $\{1,2,...,L \}$, where $L$ is the total number of categories. The goal is to partition the $n$ objects into an unknown $K$ number of clusters and identify, for each cluster, a subset of features that follow cluster-specific distributions.
Given the model parameters, each data entry is assumed to follow a categorical distribution:
\begin{equation}\label{bbc2}
\mbox{BBC2:} \  \ \  \ \ \
Y_{i,j} \mid  \bm{C},\bm{S},\bm{\Theta},K = \left\{
\begin{array}{ll}
     \text{Categorical}(\bm{\theta}_{c_i,j})  & \text{if} \quad S_{c_i,j}=1 \\
      \text{Categorical}(\bm{\theta}_{0,j})  & \text{if} \quad S_{c_i,j}=0. \\
\end{array} 
\right. 
\end{equation} 
We assume that $K-1$ follows a truncated Poisson prior with parameter $\alpha$:
\begin{align*}
\begin{split}
P(K) \propto \frac{\alpha^{K-1}}{(K-1)!}\text{exp}(-\alpha), \; \text{for} \; K=1,2,..,n.
\end{split}
\end{align*}
Conditioning on $K$, $C_i$ follows the prior distribution (\ref{prior I}) with $\gamma_0=0$ and $S_{k,j}$ follows (\ref{prior S}).  
Compared to SVD-based approaches, an attractive feature of this bi-cluster modeling approach is that its discovered features are  directly interpretable.

The binary vector $\bm{S}_j$ has $2^K$ possible configurations. However, the realization $\bm{S}_j=(1,1,\cdots,1)^T$ is indistinguishable from any of the $K$ realizations of $\bm{S}_j$ that have exactly one entry being 0 and the rest being 1s, since feature $j$ in cluster $k$ with $S_{k,j}=0$ follows a distribution not shared by any other cluster. Using $(1,1,\cdots,1)^T$ to represent the $K+1$ all cluster-specific configurations, we get a total of $2^K-K$ configurations of $\bm{S}_j$ that may lead to different likelihood values. 
 The prior for $\bm{S}_j$ conditioning on $K$ is:
\[ P(\bm{S}_j \mid K) = \left\{
\begin{array}{ll}
     \pi_S^{\sum_{k=1}^K S_{k,j}}(1-\pi_S)^{K-\sum_{k=1}^K S_{k,j}},  & \text{if} \quad \sum_{k=1}^K S_{k,j}<K\\
      \pi_S^K+K\pi_S^{K-1}(1-\pi_S),  & \text{if} \quad \text{otherwise}. \\
\end{array} 
\right. \]

With Dirichlet priors assigned to $\bm{\theta}_{k,j}, \bm{\theta}_{0 ,j}$ in (\ref{bbc2}), i.e.,
\begin{align}\label{prior theta}
\theta_{k,j}, \theta_{0 ,j}   \sim \text{Dirichlet} (\bm{\gamma}), \quad \bm{\gamma}=(\underbrace{\gamma_1, \gamma_2,...,\gamma_L}_\textrm{L})^T,
\end{align}
we integrate out $\bm{\Theta}=\{(\bm{\theta}_{0 ,j})_{j=1:p}, (\bm{\theta}_{k,j})_{{k=1:K, j=1:p}} \}$ to obtain the posterior distribution:
\begin{align}\label{Jiexing Posterior no theta}
\begin{split}
 P(\bm{C},\bm{S}, K  \mid & \mathbf{Y}) 
\propto  P(K) P(\bm{C}\mid K)P(\bm{S}\mid K)\times  \\
& 
\frac{1}{B(\bm{\gamma})^{p+\sum\limits_{j=1}^p \sum\limits_{k=1}^K S_{k,j}}} \prod_{j=1}^p \Bigg [ B \Bigg ( \sum\limits_{k: S_{k,j}=0} \bm{n_{k,j}+\gamma} \Bigg ) \cdot \prod_{k:S_{k,j}=1} B(\bm{n_{k,j}+\gamma})\Bigg ].
\end{split}
\end{align}
where  $\bm{n}_{k,j}=(n_{k,j,1},...,n_{k,j,L})^T$ with $n_{k,j,l}=\sum\limits_{i=1}^n \mathbb{I}(C_i=k, Y_{i,j}=l)$, and  $B(x_1,...,x_L)=\prod\limits_{l=1}^L\Gamma(x_l)/\Gamma(\sum\limits_{l=1}^L x_l)$.

Given $K$, the Gibbs sampler can be implemented to iteratively draw samples from the conditional posterior distributions of $C_{i}$ and $S_{k,j}$ respectively. The choice of $K$ is based on its marginal posterior probability $P(K\mid \mathbf{Y}) \propto P(K)P(\mathbf{Y} \mid K)$. To approximate the marginal likelihood $P(\mathbf{Y}\mid K)$, we consider a variant of Chib's method by making use of the posterior distribution of $\{\bm{C},\bm{S}, K\}$ in Equation (\ref{Jiexing Posterior no theta}). Specifically, we make use of the identity
\begin{align}\label{BMI_BB}
\begin{split}
P(\mathbf{Y}\mid K)=\frac{P(\mathbf{Y}\mid \bm{C},\bm{S},K)P(\bm{C}\mid K)P(\bm{S}\mid K)}{P(\bm{C}\mid \mathbf{Y},K)P(\bm{S}\mid \bm{Y},\bm{C},K)}.
\end{split}
\end{align}
All the terms in the numerator and $P(\bm{S}\mid \bm{Y},\bm{C},K)$ in the denominator can be computed exactly, while $P(\bm{C}\mid \mathbf{Y},K)$ is intractable. We can write $P(\bm{C}\mid \mathbf{Y},K)$ as:
\begin{align}\label{denom_BB}
\begin{split}
P(\bm{C}\mid \mathbf{Y},K)=\int\int P(\bm{C} \mid \bm{Y},\bm{\Theta},\bm{S},K)P(\bm{\Theta},\bm{S} \mid \bm{Y},K)d\bm{\Theta}d\bm{S}.
\end{split}
\end{align}
Similar to Section \ref{subsec: basic}, we consider the equivalent class of $(\bm{C},\bm{S})$, denoted as $(\bm{\tilde{C}},\bm{\tilde{S}})$. We have $P(\bm{C}\mid \bm{Y},K)=\frac{1}{K!}P(\bm{\tilde{C}}\mid \bm{Y},K)$. Let $(\bm{\tilde{C}}^m,\bm{\tilde{S}}^m,\bm{\tilde{\Theta}}^m)_{m=1}^M$ be outputs from the Gibbs sampler. We approximate $P(\bm{\tilde{C}}\mid \mathbf{Y},K)$ at point $\bm{C}^*$ (including its equivalent representations) by:
\begin{align*}
\begin{split}
& \hat{P}(\tilde{\bm{C}} \mid \mathbf{Y},K)=\frac{1}{M}\sum\limits_{m=1}^M P(\bm{C}^*\mid \mathbf{Y},\bm{\tilde{\Theta}}^m,\bm{\tilde{S}}^m,K).
\end{split}
\end{align*}

To demonstrate the strength of BBC2, we conduct a simulation study to compare the performances of BBC2 and a few existing methods, including STRUCTURE \citep{pritchard2000inference}, which assumes that all features follow cluster-specific distributions in all clusters, the Latent Block Model \citep{LBM14Keribin} (LBM), SUBCAD, and FABIA.
More specifically, we generate a dataset with $n=300$ objects, $p=3000$ features, $K=3$, and $\pi_S=0.15$. Parameter $\bm{\theta}$  for each categorical distribution is drawn independently  from  $\text{Dirichlet}(1,1,1)$, and the objects are equally likely to be in any cluster. The experiment is replicated 20 times independently.

When running BBC2, we let $\alpha=0.05$, $\pi_S=0.1$ and $\bm{\gamma}=(1,1)$. We start each MCMC chain at random values of $\bm{C}$ and $\bm{S}$ and run for 500 steps, with the first 200 steps as burn-in. 
The results are presented in Table \ref{tab: JX_PCA_Structure},  showing that BBC2 can correctly estimate $K$ and $\bm{C}$ in all experiments. 
In terms of its ability to infer $\bm{S}$, we calculate its feature recovery accuracy, which is the proportion of 
$\bm{S}_j$s that are correctly estimated. The feature recovery accuracy for BBC2 is as high as 91\%. Feature recovery accuracy is not applicable to the other methods. 

STRUCTURE predicts more clusters than the true number, and tends to output small clusters of sizes smaller than 10.
This overestimation is likely  caused by treating all features as following different distributions in all clusters, which brings in noise to the model. The comparison between BBC2 and STRUCTURE illustrates the importance of allowing feature homogeneity among some clusters by having a background distribution for each feature. SUBCAD also achieves 0 clustering error. However, it requires the user to input the number of clusters, which is usually unknown in advance. LBM performs slightly worse than BBC2, but is still highly accurate. 
LBM tends to overestimate the number of (row-)clusters by separating a single or a few objects from the other objects in the same underlying group. This is observed in three experiments. Computationally, to get the optimal number of (bi-)clusters, LBM requires more model comparisons than BBC2. For BBC2, We run the algorithm for a number of $K$ values. For each $K$, the algorithm automatically infers the feature matrix$\bm{S}$. However, LBM requires both the number of row-clusters and the number of column-clusters as input. Each row-cluster number and column-cluster number combination leads to a different model. 
For more than half of the datasets, FABIA cannot find any bi-cluster; and for those cases that it can find some bi-clusters, the identified clusters are of very small sizes, hence resulting in a very high clustering error rate.



\begin{table}[h!]
\begin{center}
\begin{tabular}{|l| c c c|}
\hline
Method & Number of Clusters & CE Rate & ARI\\
\hline
BBC2 & 3 & 0 & 1 \\ 
\hline
STRUCTURE & 4.9 & 5.63\% & 0.93\\
\hline
LBM & 3.25 & 0.43\% & 0.99 \\
\hline
SUBCAD & $3^*$ & 0 & 1 \\
\hline
FABIA & 1.55 & 64.1\% & 0 \\
\hline
\end{tabular}
\end{center}
\caption{Simulation results to compare BBC2 with other (Bi-)clustering methods.\\ 
*For SUBCAD, we pre-specify $K=3$.}
\label{tab: JX_PCA_Structure}
\end{table}

We also compare BBC2 with \cite{hoff2005subset}'s subspace clustering model on simulated binary outcome data (since the latter cannot be applied to other types of data). 
We run two experiments, each with 20 replicates.
In the first experiment, we fix $n=300$, $p=3000$ and choose $K=2,4,6$. In the second experiment, we fix $n=300$, $K=4$ and set $p=2000, 3000, 4000$. $\pi_S=0.25$ for all features.
Running time increases linearly with the number of features since we treat each feature independently. At $K=2$, it takes 40 seconds to finish the algorithm when $p=2000$. The time increases to 60 seconds when $p=3000$, and 80 seconds when $p=4000$. 
Since the total number of feature configurations is $O(2^{K})$, the running time increases exponentially with $K$. 
With $p=2000$ features, the running time for $K=2, 5, 7$ is 40, 120, 360 seconds, respectively. \color{black}

Clustering results for both experiments are presented in Table \ref{tab: compare with Hoff-K_p_ari}. The feature recovery accuracy and true negative rates, i.e., specificity for BBC2 and Hoff's method are shown in Table \ref{tab: compare with Hoff-S} for both experiments.
For experiment 1, 
both BBC2 and Hoff's method are robust to the increase of $K$,
 incurring zero clustering error. 
However, as shown in Table \ref{tab: compare with Hoff-S}, the feature recovery rate and true negative rate decrease with $K$ for both methods. Increasing the total number of clusters increases the chance of letting a feature to follow a cluster-specific distribution in any cluster, hence specificity decreases. Increasing $K$ also increases the total number of configurations for each $\bm{S}_j$ exponentially, resulting in more difficulties in fully recovering $\bm{S}_j$.
 However, BBC2 maintains a higher feature recovery accuracy (by 50\% or more) and a higher true negative rate (by 30\% or more) than Hoff's method in all $K$ values. 
 When we increase the total number of features in experiment 2, performances for both methods stay unchanged with both retaining the zero clustering error rate. The feature recovery rate stays at 69\% for BBC2, and 45\% for Hoff's. BBC2 also maintains a much higher true negative rate (91\%) than Hoff's method (51\%).

SUBCAD also achieves zero clustering error in all settings, and  it depends on knowing the true number of clusters upfront. LBM overestimates the number of clusters. The clustering error rate increases with $K$s. In many cases, it outputs small clusters having only a few objects. There are also cases that objects from the same underlying cluster are split into two clusters of roughly equal size, and cases that two underlying clusters are estimated to be one. These cases can significantly increase the error rate. For experiment 1, FABIA can not find any bi-cluster when $K=2$, and has a much lower ARI than other methods. 
Its performance is also quite variable: in some datasets, it can recover the underlying clustering patterns, but in others, it incurs significant clustering errors. 

\begin{table}[H]
\begin{center}
\begin{tabular}{|l | c|c c c|c| c c c |}
\hline
 Method & Vary K, & Number of & CE & ARI & Vary p, & Number of & CE & ARI\\
& Fix p & Clusters & Rate &  & Fix K & Clusters & Rate &\\
\hline
BBC2 & &  2& 0 & 1 & & 4 & 0 & 1\\ 
\cline{1-1}\cline{3-5}\cline{7-9}
Hoff & K=2 &  2 & 0 & 1 & p=2000 & 4 & 0 & 1 \\ 
\cline{1-1}\cline{3-5}\cline{7-9}
LBM & (p=3000) &  2.6 & 0.83\% & 0.98 & (K=4) & 4.95 & 1.37\% & 0.98\\
\cline{1-1}\cline{3-5}\cline{7-9}
SUBCAD & &  $2^*$ & 0 & 1 & & $4^*$ & 0 & 1 \\ 
\cline{1-1}\cline{3-5}\cline{7-9}
FABIA & &  1 & 47.47\% & 0 & & 3.65 & 20.68\% & 0.69  \\
\hline
\hline
BBC2 &&  4& 0 & 1 & & 4 & 0 & 1\\
\cline{1-1}\cline{3-5}\cline{7-9}
 Hoff & K=4 & 4& 0 & 1 & p=3000 &4 & 0 & 1\\ 
\cline{1-1}\cline{3-5}\cline{7-9}
LBM & (p = 3000) & 5.45 & 3.15\% & 0.97 & (K=4) & 4.7 & 4.97\% & 0.94\\
\cline{1-1}\cline{3-5}\cline{7-9}
SUBCAD &  & $4^*$ & 0 & 1 & & $4^*$ & 0 & 1\\
\cline{1-1}\cline{3-5}\cline{7-9}
FABIA & & 2.55 & 22.47\% & 0.67 & & 3.8 & 18.6\% & 0.70 \\
\hline
\hline
BBC2 & &  6&  0 & 1 & &4 & 0 & 1\\
\cline{1-1}\cline{3-5}\cline{7-9}
Hoff & K=6 & 6& 0 & 1 & p=4000 &4 & 0 & 1\\
\cline{1-1}\cline{3-5}\cline{7-9}
LBM &(p = 3000)  & 7.7 & 10.85\% & 0.89 & (K=4) & 6.2 & 2.97\% & 0.96\\
\cline{1-1}\cline{3-5}\cline{7-9}
SUBCAD & & $6^*$ & 0 & 1 & & $4^*$ & 0 & 1\\
\cline{1-1}\cline{3-5}\cline{7-9}
FABIA & & 3.45 & 27.88\% & 0.65 & &  3.4 & 21.2\% & 0.69\\
\hline
\end{tabular}
\end{center}
\caption{Comparing BBC2 with other Bi-clustering methods for binary data under different simulation settings.\\
*For SUBCAD, we input the true number of clusters for all experiments.}
\label{tab: compare with Hoff-K_p_ari}
\end{table}

\begin{table}[h!]
\begin{center}
\begin{tabular}{|l | c c c| c c c|}
\hline
Simulation&  \multicolumn{3}{c|}{Change number of clusters K} & \multicolumn{3}{c|}{Change number of features p}\\
Settings  &  \multicolumn{3}{c|}{(p=3000) } &\multicolumn{3}{c|}{(K=4) }\\
\hline
\diagbox{Method}{K(p)} & K=2 & K=4 & K=6 & p=2000 & p=3000 & p=4000\\
\hline
 BBC2 & 86.3 / 96.5 & 68.8 / 91.0 & 57.7 / 85.6 & 69.0 / 91.0 & 68.8 / 90.9 & 68.6 / 91.0 \\
\hline
 Hoff & 50.5 / 73.3 & 45.3 / 52.0 & 33.6 / 32.9 & 45.5 / 51.6 & 45.5 / 50.8 & 45.2 / 51.0\\ 
\hline
\end{tabular}
\end{center}
\caption{Feature recovery accuracy (RA) and true negative rate (TNR) of BBC2 and Hoff's method.
Numbers in each cell are in percentage, with the first representing the RA and the second  the TNR.}
\label{tab: compare with Hoff-S}
\end{table}

\subsection{Hierarchical Bayesian bi-clustering}\label{subsec: hbbc}
Philosophically, the task of ``clustering'' (aka {\it unsupervised learning}) is a very subjective matter and can result in drastically different findings if one focuses on different things, or even different levels of details. For example, given a group of people, one may cluster them into rough ethnicity groups, e.g., Asian, Caucasian, African, etc. However, further dividing them into a finer level of ethnicity groups (e.g., Chinese, Japanese, etc) will need some professional training. Indeed, as shown in the genetics example in Section \ref{application: HapMap}, BBC2 can well separate distant populations, but tends to group intra-continent sub-populations into the same cluster because it has a hard time re-focusing its attention to more refined details. This observation argues that a unified clustering model might not be able to simultaneously account for both large and subtle differences. A natural solution similar to human's learning strategy   is to do clustering recursively or hierarchically \citep{upgma,johnson1967hierarchical}.
As attempted in \cite{Jiexing}, we extend the classical hierarchical clustering  strategy to ``hierarchical Bayesian bi-clustering'' (HBBC, henceforth), in which  we build a tree based on BBC2. As the tree grows,one leaf node is allowed to be further split into two child nodes, and the choice of whether and which leaf node to split  are governed by a Bayesian criterion.

HBBC starts with assigning all objects to one cluster - the root node (step 0). Each subsequent step splits an existing cluster and increases the number of clusters by one until the tree stops growing. At the beginning of step $t$, $t$ clusters have been formed with corresponding data blocks $\{\bm{Y}^r \}_{r=1}^t$, a row-wise partition of the whole dataset $\bm{Y}$. We assume that the data of all clusters are mutually independent conditioning on the corresponding hierarchical structure. Denote the current hierarchical structure as model $H_0$, and its ``\textit{descendant}'' structure as $H_r$ ($r=1,2,\cdots,t$) if node-splitting takes place at 
leaf node $r$. Note that $H_0$ corresponds to the case that the tree terminates at step $t$. We assign a prior probability of $q/t$ to each $H_r$ model, and $1-q$ to model $H_0$, where $0<q<1$ is a hyper-parameter. As the tree grows deeper, it is harder to split further at any leaf node {\it a priori}. 

To determine whether and where the tree should get split, we consider the potential bi-clustering for data block $\bm{Y}^r$ at step $t$. Let $K_r$ be the number of clusters it can split into. Under $H_0$, $K_r=1$, and under $H_r$, $K_r=2$. The model for $\bY^r$ and the priors for $K_r$, $\bm{C}^r$ and $\bm{S}^r$ all follow BBC2's setup.
We define $w_r$ to be a step-scaled ratio of model posterior probabilities under $H_r$ and $H_0$:
\begin{align*}
    \begin{split}
        w_r :&\equiv t\cdot \frac{P(H_r\mid \bm{Y})}{P(H_0 \mid \bm{Y})}= \frac{q}{1-q}\frac{P(\bm{Y}\mid H_r)}{P(\bm{Y}\mid H_0)}=\frac{q}{1-q} \frac{\prod_{l=1}^t P(\bm{Y}^l\mid H_r)}{\prod_{l=1}^t P(\bm{Y}^l\mid H_0)}\\
        &=\frac{q}{1-q} \frac{ P(\bm{Y}^r\mid H_r)}{P(\bm{Y}^r\mid H_0)} = \frac{q}{1-q} \frac{ P(\bm{Y}^r\mid K_r=2)}{P(\bm{Y}^r\mid K_r=1)} ,
    \end{split}
\end{align*}
and $w_0=t$ by default.

To calculate $w_r$, $r=1,\cdots, t$, the only difficult part is $P(\bY_r \mid K_r=2)$, which corresponds to Equation (\ref{BMI_BB}). It has been dealt with in the previous subsection with a variant of  Chib's method. $P(\bY_r \mid K_r=1)$ corresponds to the case that all objects are the same cluster. Based on the model defined in (\ref{bbc2}) and (\ref{prior theta}), it can be easily computed as
\begin{align*}
    \begin{split}
        P(\bY^r \mid K_r=1) &=\int P(\bY^r \mid \bm{\Theta}, K_r=1)P(\bm{\Theta} \mid K_r=1) d\bm{\Theta} \\
        &= \prod\limits_{j=1}^p P(\bY^r_{\cdot,j} \mid \btheta_j, K_r=1)P(\btheta_j \mid K_r=1)d\btheta_j = \prod\limits_{j=1}^p \frac{B(\bm{n_{r,j}}+\bm{\gamma})}{B(\bm{\gamma})}.
    \end{split}
\end{align*}

At each step, we identify the largest $w_r$, $r=1,\cdots,t$. If it is greater than $w_0=t$, we split the corresponding node, and go to the next step. Otherwise, the tree stops at level $t$.

\subsection{Bayesian bi-clustering for  data integration}\label{Integrative}
Although some Normal distribution-based Bayesian bi-clustering methods have been developed earlier as reviewed in Section~\ref{intro}, they are  mostly used for analyzing  ``homogeneous'' data types (such as gene expression data produced by the same technology from the same lab) and are not flexible  enough to handle complex data structures, such as those encountered in integrative genomics tasks.
For example, a typical integrative genomic problem is to combine many thousands of published mRNA  expression datasets (such as those in GEO and TCGA) to gain deeper biological insights and suggest new biological discoveries. 
It is inappropriate to impose a simple Bayesian bi-clustering structure developed in Section \ref{sec: general framework} to the combined dataset since different labs and technologies typically induce different types of error distributions and normalization issues. Instead, one can work on the gene correlation matrices resulting from different datasets, which can be regarded as data summaries and are more uniform and comparable with each other. A correlation matrix can also be a reasonable  summary of discrete data, especially in high dimensions.
The ``CLIC'' algorithm developed in \cite{li2017clic} is based on this observation, of which we here describe a generalized formulation.

Let $\bm{X}_1,\ldots, \bm{X}_p$ denote $p$  datasets (e.g., microarrays), each with $N$ rows (corresponding to $N$ genes), and let $\bm{X}_d$ have $l_d$ columns (corresponding to $l_d$ experimental samples). For robustness, we can screen out those datasets with $l_d<10$. We first convert dataset $\bm{X}_d$ into a transformed $N\times N$ gene correlation matrix, $\bY_d$, whose $(i,j)$-th entry is $Y_{d,i,j}={1\over 2} \log \frac{1+r_{d,i,j}}{1-r_{d,i,j}}$, with $r_{d,i,j}$ being the sample correlation coefficient between gene $i$ and $j$ in dataset $d$. Suppose we have a query list $Q_n$ of $n$ genes (a subset of the $N$ genes), and are interested in inferring gene co-expression modules, i.e., ``clusters,'' among genes in $Q_n$ and datasets in which these modules are active. Each co-expression module with the relevant datasets forms a bi-cluster. 
After identifying these bi-clusters, we may fish out more genes not in $Q_n$ for each  module  based on their co-expression patterns  with the inferred module in the selected datasets. 

Again, we let $\bm{C}=(C_1,\ldots, C_n)^T$ be the vector of cluster assignment as before, with each $C_i$ taking values in $\{0,1,\cdots,K\}$, with $k=0$ representing a null cluster.
Each non-null cluster $k$ is assumed to be active in only a subset of datasets. Let $S_{k,d} \in \left\{0,1\right\}$ indicates whether dataset $d$ is selected or not for cluster $k$. If dataset $d$ is selected into cluster $k$, genes in cluster $k$ co-express over the set of experimental conditions in dataset $d$, and we assume their transformed sample correlations follow the same (cluster-specific) normal distribution with mean $\theta_{k,d}$ and variance $\sigma_{k,d}^2$. If dataset $d$ is not selected ($S_{k,d}=0$) or genes $i$ and $j$ are not in the same non-null cluster, the corresponding $Y_{d,i,j}$ is normally distributed with dataset-specific mean $\theta_{0,d}$ and variance $\sigma_{0,d}^2$. In summary, for $k=1,\dots,K$,
\begin{align*}
[Y_{d,i,j}\mid S_{k,d}=1,C_{i}=C_{j}=k] & \sim N\left(\theta_{k,d},\sigma_{k,d}^{2}\right),\\
[Y_{d,i,j}\mid S_{k,d}=0\text{ or }C_{i}\neq C_{j}\text{ or }C_i=C_j=0 ] & \sim N\left(\theta_{0,d},\sigma_{0,d}^{2}\right).
\end{align*}

For computational ease, we adopt the
conjugate Normal-Inv-Gamma prior for $(\theta_{k,d}, \sigma_{d,k}^2)$. Priors for $\bm{C}$ and $S_{k,d}$ follow (\ref{prior I}) and (\ref{prior S}). Since the number of genes of a typical gene module is no more than a few hundreds, in most practical cases, the number of gene pairs in which both genes are within the same co-expression module is very small compared to the total number of gene pairs $N(N-1)/2$, we thus fix the dataset-specific background parameters $\theta_{0,d}$ and $\sigma_{0,d}^2$ for two random genes not in the same module at their total-sample estimates for dataset $d$, i.e.,
\begin{align*}
\theta_{0,d}=\frac{2}{N\left(N-1\right)}\sum_{1\le i<j\le N}Y_{d,i,j},\quad\sigma_{0,d}^{2}=\frac{2}{N\left(N-1\right)}\sum_{1\le i<j\le N}\left(Y_{d,i,j}-\theta_{0,d}\right)^{2}.
\end{align*}
Furthermore, we can integrate out $\btheta$, $\bm{\sigma}$ and $\bm{S}$ to obtain the marginal likelihood $P(\bm{Y} \mid \bm{C}, K)$ and implement a Gibbs sampler on the space of  $(\bm{C}, K)$ (see Appendix \ref{app: clic} for details).

In the above formulation, although bi-clusters correspond to gene-dataset combinations, the actual data we work with are matrices of the transformed gene pairwise correlations
obtained from different datasets. To connect  back to our general framework, we can reformat the $p\times n \times n$ array $\bY$ to a $n(n-1)/2 \times p$ matrix $\tilde{\bm{Y}}$, where each row corresponds to a pair of query genes (recording their correlation), and each column a dataset. Then, we can map the clustering of the original set of genes to the clustering of the rows of $\tilde{\bm{Y}}$: 
a row of $\tilde{\bm{Y}}$  is assigned to cluster $k \in \{1,\ldots, K \}$ if the two genes correspond to this row 
are both in cluster $k$, otherwise the row is assigned to cluster 0. In other words, this formulation differs from the typical bi-clustering settings in that  the clustering of the rows of $\tilde{\bm{Y}}$ has an additional restrictive structure, which can be reflected in its prior.
The rest follows the general framework of Section \ref{sec: general framework}.

What we just described can be viewed as the second stage of a two-stage procedure. In the first stage we obtain a summary for each dataset in the form of a correlation matrix of the objects involved --  the individual features of the objects are no longer used in the second stage although the  derivation of this summary matrix relies on the features and may involve feature selections. Our bi-clustering model is designed only for the second stage, i.e., the bi-clustering of objects and datasets. 
 Generalized Biclustering (GBC) developed in \cite{GBC2020} is also designed to integrate information from data of different types. GBC concatenates  all data matrices that share a set of common objects, retaining all features without any data reduction. It models this concatenated matrix  jointly through an exponential family-based factor model and conducts bi-clustering on all the features and all the objects. While GBC is more delicate and principled, our approach is more robust and much more scalable (e.g., can be applied to combine information from thousands of datasets each with hundreds to many thousands features).
In principle,  our method  can be applied to integrate datasets of various types, provided that information of each dataset can be approximated summarized as a  correlation matrix for certain characteristics (e.g., mRNA expressions, or mutation loads)  of the objects (e.g.,  genes) common to all datasets.

\section{Real-data applications of Bayesian bi-clustering}\label{application: HapMap}
%
\subsection{Bi-clustering analysis of genetic data by BBC2}\label{sec: bbc2 hapmap}
Let us reconsider the human genetic data introduced in Section~\ref{sec:case}. BBC2 assumes that features are independent of each other. To obtain a set of SNPs that are not in linkage disequilibrium (i.e. nearly independent), we use PLINK to remove correlated SNPs by setting the threshold for the squared correlation $r^2$ to be $10^{-6}$ over a sliding window of 200 variants and an increasing step of 20.  We have also tried other thresholds such  as $10^{-4}$ and $10^{-2}$,
and find the clustering result relatively robust (see Appendix \ref{app: Hapmap_r2} for details). All children  from trio-families are dropped, resulting in unrelated individuals. The two filtering steps lead to a dataset of 4,217 SNPs and 1,198 individuals from 11 populations based on population origin. A description of the dataset can be found in Appendix \ref{app: HapMap Shuffled}.


To run BBC2, we let $\alpha=0.05$, $\beta=0.05$, and $\bm{\gamma}=(1,1,1)$. A small $\alpha$ represents a prior belief that samples come from a small number of clusters. A small $\beta$ expresses the prior belief that most features are background ones shared across clusters. The hyper-parameter of the Dirichlet prior for the genotype frequency $\bm{\theta}$ of each SNP  is set at $\bm{\gamma}=(1,1,1)$.
The total number of MCMC steps is 600 with the first 200 steps as burn-in.\color{black} 

BBC2's clustering result is shown in Table \ref{tab: HapMap Result}, with an ARI of 0.56. The method estimates the number of clusters to be 6. Individuals from the ASW population are split into two clusters, mixing with the other populations with African ancestry. For the rest of the populations, almost all individuals from the same population are grouped together. Populations in mixed groups are close to each other geographically, i.e., from the same continent. Out of the 4,217 SNPs, 96.42\% show certain extent of diversity cross populations. 
Table \ref{tab: SNP overlap} summarizes the number of SNPs that follow cluster-specific distributions in the specified numbers of clusters. For example, only 151 SNPs are background features following a common distribution,
and only 11 SNPs are all-cluster-specific, meaning that their distributions are different in different clusters. These numbers suggest neither clustering nor clustering with feature selection is adequate in modeling the heterogeneity in the distributions of features.

When we specify the number of clusters to be 11, the true number of populations, BBC2 still outputs 6 nonempty clusters, and the cluster assignments for the 1,198 individuals are exactly the same as Table \ref{tab: HapMap Result}. 
This suggests that more refined (weak) signals do not seem to help in global clustering.
To further investigate this phenomenon, we randomly select $R\%$ of the features (SNPs) and randomly permute their observations (so that these features are made to be noninformative). We rerun the algorithm with $R$ ranging from 10 to 90. Clustering error rate stays almost the same when $R$ is below 75, but  increases to about 54\% after that. This observation indicates that a majority of the SNPs carry some amount of information, 
which supports the previous finding that over 95\% of the SNPs are ``important''. It also indicates BBC2 is robust to the presence of many noninformative features. Among the randomized features, the algorithm correctly estimates 90-97\% of them to follow background distributions, i.e., 3-10\% false positive rate.
More details can be found in Appendix \ref{app: HapMap Shuffled}.


\begin{table}[h!]
\begin{center}
\begin{tabular}{|c |c c c c c c|}
\hline
Population & Cluster 1 & Cluster 2 & Cluster 3 & Cluster 4 & Cluster 5 & Cluster 6 \\
Abbreviation & & & & & & \\
\hline
\hline
ASW & 31& 22 & 0& 0& 0& 0\\
\hline
LWK & 110& 0& 0& 0& 0& 0\\
\hline
MKK & 4& 152& 0& 0& 0& 0\\
\hline
YRI & 147& 0& 0& 0& 0& 0\\
\hline
GIH & 0& 0& 101& 0& 0& 0\\
\hline
MEX & 0& 0& 0& 53& 5& 0\\
\hline
CHB & 0& 0& 0& 0& 0& 137\\
\hline
CHD & 0& 0& 0& 0& 0& 109\\
\hline
JPT &0& 0& 0& 0& 0& 113\\
\hline
CEU &0 &0 &0 &0 &112 & 0\\
\hline
TSI &0 &0 &0 & 0&102 & 0\\
\hline
\end{tabular}
\end{center}
\caption{BBC2's clustering result on the human genetic dataset: rows correspond to population origins and columns correspond to estimated clusters.}
\label{tab: HapMap Result}
\end{table}


\begin{table}[h!]
\begin{center}
\begin{tabular}{|l|c c c c c c|}
\hline
Number of Cluster-Specific Distributions &0 & 1 & 2 & 3 &4 &6 \\
\hline
Number of SNPs & 151 & 611 & 1448 & 1553 & 443 & 11 \\
\hline
\end{tabular}
\end{center}
\caption{Numbers of SNPs (second row) that are estimated by BBC2 to have the specified numbers (first row) of cluster-specific distributions for the human genetic dataset. }
\label{tab: SNP overlap}
\end{table}

\subsection{Analyzing the genetic data with HBBC}\label{sec: hbbc hapmap}
In the previous subsection, 
the grouping of intra-continent sub-populations into the same cluster by BBC2 indicates that intra-continent minor variations across sub-populations have been missed out. We demonstrate here that HBBC, with its hierarchical nature, is able to pick up these weak signals.
From the modeling perspective, focusing on one node a time enables us to detect minor differences between closely related sub-populations. It can also capture the scenario that a feature has several background distributions, corresponding to different higher level groups, e.g., continents. The feature can have sub-population-specific distributions in certain continents, and continent-specific background distributions in the rest. From the evolutionary perspective, human population 
 evolves roughly in a hierarchical fashion (the out-of-Africa theory): Homo sapiens developed first in Africa and then spread to other continents about 50-70,000 years ago \citep{OOA_Theory_2, OOA_Theory_posth2016}. Then, within a continent, sub-populations emerged at different times. Hence, sub-populations within a continent are genetically closer to each other, and are on a smaller sub-tree before joining other sub-populations in other continents to form a larger (sub)-tree.

\begin{figure}[H]
\centering
\includegraphics[width=0.8\linewidth, height=2.5in]{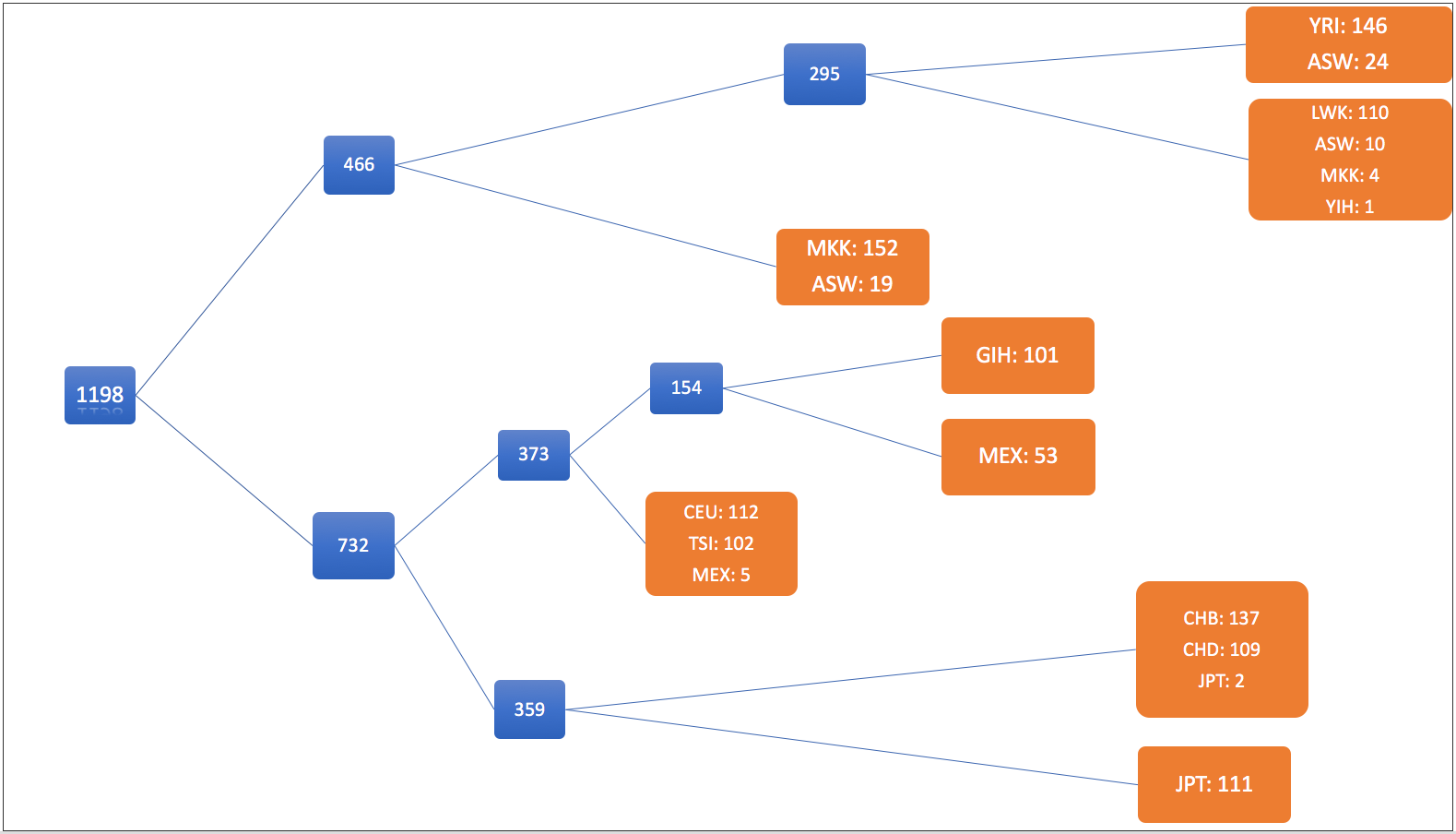}
\caption{HBBC's estimation result on the human genetic data, minimum node size=50. The number in each blue node is the number of individuals assigned to the node. In each orange node, i.e., a leaf node: each population name-number pair indicates for that population, the number of individuals that are assigned to this node. Reading from the left to the right of the figure gives the order of the node splitting process.}
\label{fig:SNP_tree}
\end{figure}

We apply HBBC with hyper-parameter $q=0.05$, reflecting a fairly conservative prior on developing new clusters. To further alleviate excessive learning, we restrict the minimum size of any node to be 50, which corresponds to around 5\% of the total number of individuals. We have also tested different minimum node sizes: 5, 20, 30 and 40, and found the results to be robust (see Appendix \ref{app: min node size} for details). Other parameters are set to the same values as in BBC2. 
The result is shown in Figure \ref{fig:SNP_tree}, with the leaf nodes colored in orange. Each blue internal node is labelled with the number of individuals assigned to it. The numbers inside each orange node are numbers of individuals by population. 
At the first split, populations with African ancestry are  separated from the rest; at the second split, the East Asians are separated from individuals with European  (CEU, TSI) or European-mixed (MEX, GIH) ancestries\footnote{Some recent studies show that there is a strong European influence on the genetic composition of the Indian population \citep{IndoEuro_reich09, IndoEuro_reich19}. \cite{euro_amer} and \cite{mex_admixture} also document the genetic proximity of European populations and the native American, and Latin American populations.}, followed by the third and fifth splits that further separate populations in the latter three ancestry groups. The sixth step identifies subpopulations in the East Asian group: the JPT samples are separated from the two Chinese populations. 
Among the populations with African ancestry, the MKK population and a small fraction of the ASW population are the first to be separated from the rest. The second major cluster with African ancestry consists of the YIH population and another subset of the ASW population. The LWK population makes up the third major cluster. The ARI of the clustering result is equal to 0.75.

The advantage of HBBC over BBC2 has been demonstrated in this example through its ability to distinguish the Japanese from the Chinese, and the Luhya samples from the Yoruba samples. The order of the node splitting process is likely to imply the relative genetic proximities of different populations. For example, each of the first three steps separates populations by continents. While intra-continent populations are identified after step 3.
In terms of selected features, the number of SNPs that have different distributions between the Chinese and the Japanese populations is estimated to be 354 (8.39\%), while this number is 1,053 (24.97\%) between the GIH population and the MEX population. Geographically, GIH and MEX are more distant to each other than JPT and CHB+CHD. Among the African populations, the MKK-dominate group is most different from the rest of the African populations as it is the first to be split out and is distinguished on 1,167 (27.67\%) SNPs, while the number is only 317 (7.52\%) between the YIH-dominate group and LWK-dominate group.

Benefits of identifying a set of SNPs that have different distributions in different ethnicity groups are not limited to clustering of individuals.
For example, if we want to estimate polygenic risk scores of a certain disease for a group of individuals, the developed bi-clustering methods (BBC2 and HBBC) help account for hidden population structures and estimate the baseline distributions of SNPs more accurately, which may result in a better risk score estimate. 
The SNPs selected in each step of HBBC are also potentially related to human phenotypes that show variations among different populations. We take human hair and hair color related traits as an example. The differences in the traits are larger for populations in different ethnicity groups across continents, while it is more subtle when comparing geographically closer populations. So we should expect to see the traits-related SNPs becoming less likely to be selected in later steps of HBBC. The analysis of hair-related SNPs in Appendix~\ref{sec:hair} provides some support.

\subsection{Clustering single-cell RNA sequencing data using BBC1}\label{subsec: scRNAseq}
We apply BBC1 to the dataset in \cite{chu2016single}. The dataset contains the expression levels of 19,097 genes in 1,018 cell samples from 7 human cell types: DE cells (endoderm derivatives), endothelial cells (EC), neuronal progenitor cells (NPC), trophoblast-like cells, single undifferentiated H1 and H9 ES cells, and foreskin fibroblasts (HFF). We binarize the data such that 1 represents a high expression level and 0 otherwise before applying BBC1. The prior probability of a gene being a biomarker is set to be $\pi_S=0.1$. Flat priors are given to $\theta$s: i.e., $\alpha_{\theta_1}=\alpha_{\theta_2}=\alpha_{w_1}=\alpha_{w_2}=1$. The number of MCMC steps is 700 with the first 200 as burn-in. We compare BBC1's clustering result to those produced by SC3 \citep{sc3} and Seurat \citep{Seurat}. The original, un-binarized, data are used for both SC3 and Seurat. For Seurat, 20 principal components are used for its \textit{FindNeighbors} function and the resolution is set to 0.5 for the function \textit{FindClusters}.

The clustering results from the three algorithms are shown in Tables \ref{tab: chu-BBC1}-\ref{tab: chu-sc3}. SC3 gives the highest ARI of 0.93. BBC1 is the second best with an ARI of 0.74, closely followed by Seurat with an ARI of 0.73. BBC1 and Seurat have very similar clustering results, both having difficulties in separating H1 and H9 cells. Samples from these two cell types are mixed together over the first 2 principal component directions. The higher ARI of SC3 is entirely attributed to its ability to distinguish H1 and H9 cells. SC3 performs slightly worse in clustering  DE cells compared to BBC1 and Seurat. Though all algorithms separate NPC into two clusters, BBC1 is able to put the majority of NPCs into one cluster, and the rest to a minor cluster. SC3 and Seurat tend to separate NPCs into two equally sized clusters. BBC1 also put all EC  in one cluster (i.e., Cluster 2 in Table \ref{tab: chu-BBC1}), while SC3 and Seurat both have one EC  mixed with DE cells. 
There are 9,837 genes whose   posterior probabilities of being biomarkers is greater than 0.99 according to BBC1. The number of marker genes predicted by SC3 is 8,499, and 5,548 by Seurat. BBC1 picks up 87\%  and 51\%  of the marker genes from SC3 and Seurat, respectively, as biomarkers. 

Comparing the performance of BBC1 with SC3 and Seurat, we observe that binarization does not  result in much loss of information, but rather reduces noise and improves robustness. In  certain extremely noisy cases and/or when model assumptions are severely violated, we believe that BBC1 is more likely to cluster samples from the same cell type together than existing approaches. 

\begin{table}[H]
\begin{center}
\begin{tabular}{|l| c| c| c| c| c| c| c|}
\hline
Cell Type & Cluster 1 & Cluster 2 & Cluster 3 & Cluster 4 & Cluster 5 & Cluster 6 & Cluster 7\\
\hline
\hline
DEC  & 138  & 0 &0 &   0  & 0&   0  & 0\\
\hline
EC    & 0 &105&  0&   0&   0 &  0 &  0 \\
\hline
H1 & 0  & 0  &207  &  0  & 0   &0  & 5\\
\hline
H9  &  0 &  0& 162 &  0&   0&   0  & 0\\
\hline
HFF &  0 &  0  & 0& 0 & 159  & 0 &  0\\
\hline
NPC   &0  & 0 &  0  & 43 & 0& 130 &  0\\
\hline
TB  &  0 &  0   &0   &0  & 0 & 0 &69\\
\hline
\end{tabular}
\end{center}
\caption{BBC1's estimation result on the scRNA-seq data. Rows correspond to true cell types, and columns correspond to estimated clusters.}
\label{tab: chu-BBC1}
\end{table}

\begin{table}[H]
\begin{center}
\begin{tabular}{|l| c| c| c| c| c| c| c|}
\hline
Cell Type & Cluster 1 & Cluster 2 & Cluster 3 & Cluster 4 & Cluster 5 & Cluster 6 & Cluster 7\\
\hline
\hline
DEC & 137&   1&   0 &  0&  0   &0 &  0\\
\hline
EC & 1& 104 &    0 &  0  & 0 &  0  & 0\\
\hline
H1 & 0  & 0 & 212 &  0   &  0 &  0 &  0\\
\hline
H9 &  0 &  0& 162  & 0  &  0  & 0  & 0\\
\hline
HFF  &0  & 0  & 0& 0  &  159 &  0 &  0\\
\hline
NPC &  0 &  0  & 0 &  79 &  0 & 94 &  0\\
\hline
TB   & 0&   0   &0 &  0  & 0  & 0  &69\\
\hline
\end{tabular}
\end{center}
\caption{Seurat's clustering result on the scRNA-seq data. Rows correspond to true cell types, and columns correspond to estimated clusters.}
\label{tab: chu-Seurat}
\end{table}

\begin{table}[H]
\begin{center}
\begin{tabular}{|l| c| c| c| c| c| c| c|c|c|}
\hline
Cell  & Cluster & Cluster & Cluster & Cluster & Cluster & Cluster & Cluster & Cluster & Cluster\\
Type & 1 & 2 & 3 &4 & 5 & 6 & 7 & 8 & 9\\
\hline
\hline
DEC &  127 &  0 & 0  &  0 & 0  & 0 &  0& 0&11  \\
\hline
EC  &  0 & 104 &  0&  0&    0&   0 &  0  &  0&  1\\
\hline
H1  &  0 &  0 &212 &  0  & 0 &  0 &  0 &  0  & 0\\
\hline
H9   & 0 &  0 & 0 &162  & 0  & 0 &  0 &  0 &  0 \\
\hline
HFF &0 & 0   &0  &  0  & 159&  0    &0  & 0 &  0\\
\hline
NPC &  0 &  0 & 0 & 0  & 0 & 89  & 0 &   84  & 0 \\
\hline
TB  & 0  &  0&   0 &  0&  0 &  0  & 69  & 0 & 0 \\
\hline
\end{tabular}
\end{center}
\caption{SC3's clustering result on the scRNA-seq data. Rows correspond to true cell types, and columns correspond to estimated clusters.}
\label{tab: chu-sc3}
\end{table}

\color{black}

\subsection{Gene clustering  based on co-expression patterns}
To demonstrate the bi-clustering approach described in Section \ref{Integrative} for integrative genomics, we amass a collection of 1,774 mRNA datasets from GEO for the mouse Affymetrix chip Mouse430\_v2 platform \citep{li2017clic}, and apply the correlation-matrix-based bi-clustering algorithm to a list of 58 genes. The ground truth, which is unknown to the algorithm, is that the genes are from two pathways:  the  DNA replication pathway with 22 genes, and the mitochondrial respiratory chain complex I with 36 genes. 

\begin{figure}
\centering
\includegraphics[width=6in,height=5in]{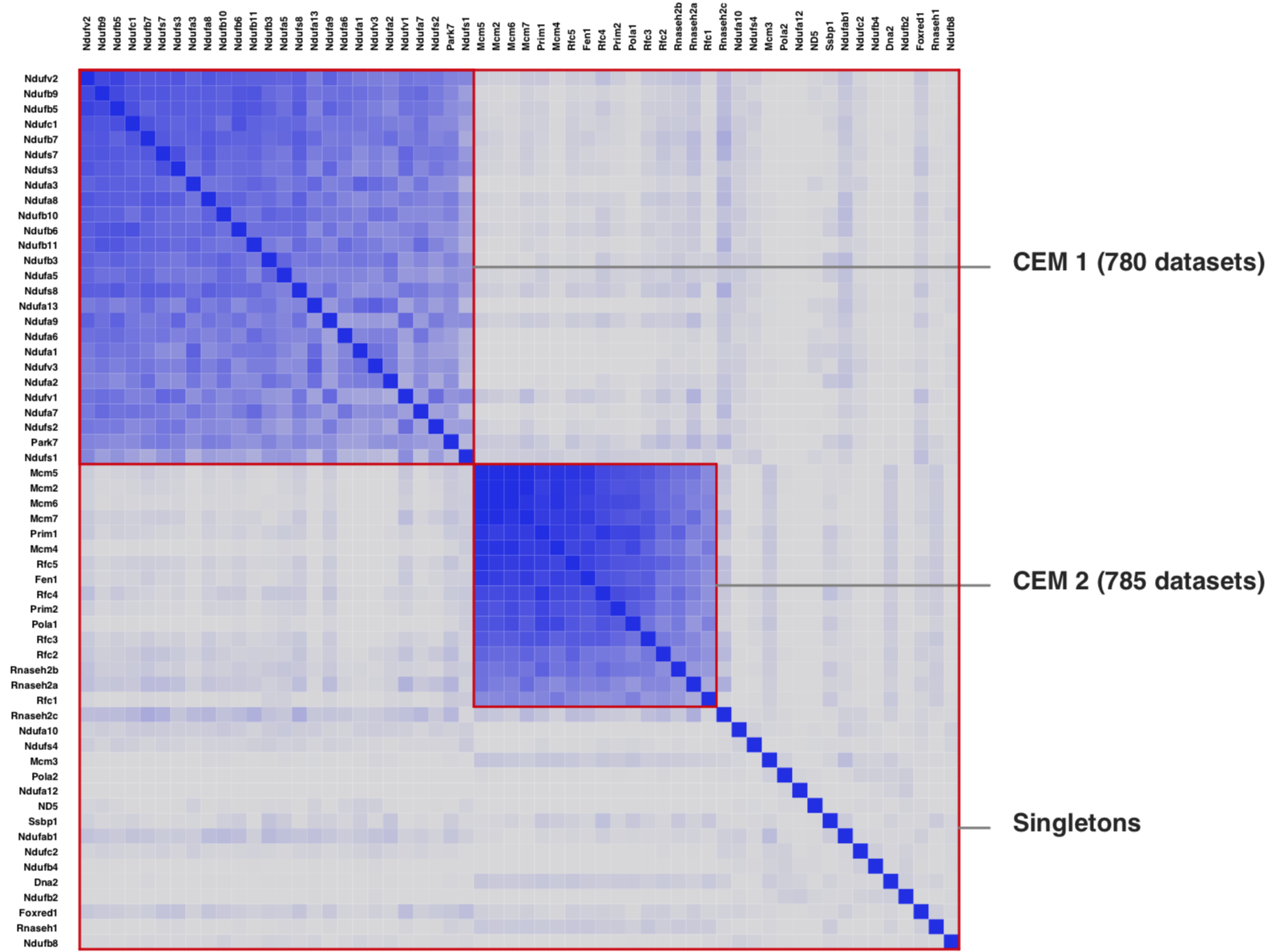} 
\caption{CLIC's clustering result on the input two pathways. Both rows and columns correspond to the input genes. CEM1 consists of 26 genes from one pathway and is supported by 780 datasets. CEM2 consists of 16 genes from the other pathway and is supported by 785 datasets. Genes that are in neither CEM are identified as singletons.}
\label{fig: CLIC_2CEM}
\end{figure}

Figure \ref{fig: CLIC_2CEM} displays the clustering result, indicating that the algorithm identifies two nontrivial gene co-expression clusters, CEM1 and CEM2. CEM1 consists of 26 genes from the pathway for mitochondrial respiratory chain complex I (missing 10 genes) and CEM2 consists of 16 genes from the DNA replication pathway (missing 6 genes). No gene from one pathway is assigned to the cluster of the other pathway. The 16 genes not included in the two clusters are identified as singletons.

Genes in CEM1 are estimated to be strongly co-expressed in 780 datasets, wheases those in CEM2 are supported by 785 datasets. For datasets that support both clusters, different co-expression patterns can occur in the two clusters. As an illustration, we select one such dataset with 59 experimental samples, and show its gene expression patterns in Figure \ref{fig: CEM_Partial}. In the first 19 samples, genes in CEM1 are all highly expressed, while all genes in CEM2 have low expression values. The pattern is reversed in the next 21 samples. In the last 19 samples, genes in CEM1 show low expression levels, while moderate expression levels are observed among genes in CEM2. The different co-expression patterns across samples in one dataset for one cluster is due to the fact that different samples correspond to different experimental conditions.
Analyzing the expression patterns in the two clusters across datasets tells us when the two pathways will be both activated and work together.
Examining the selected datasets  helps characterize conditions under which genes in the cluster co-express and reveals
insights on biological functions of the gene clusters.
For example, among the top 10 datasets selected by CEM1, 5 of them are related to inflammatory processes. Two are related to experiments studying rheumatoid arthritis and another two are related to inflammation in liver. 

\afterpage{
\begin{landscape}
\begin{figure}
\includegraphics[width=1.6\textwidth, height=0.9\textheight]{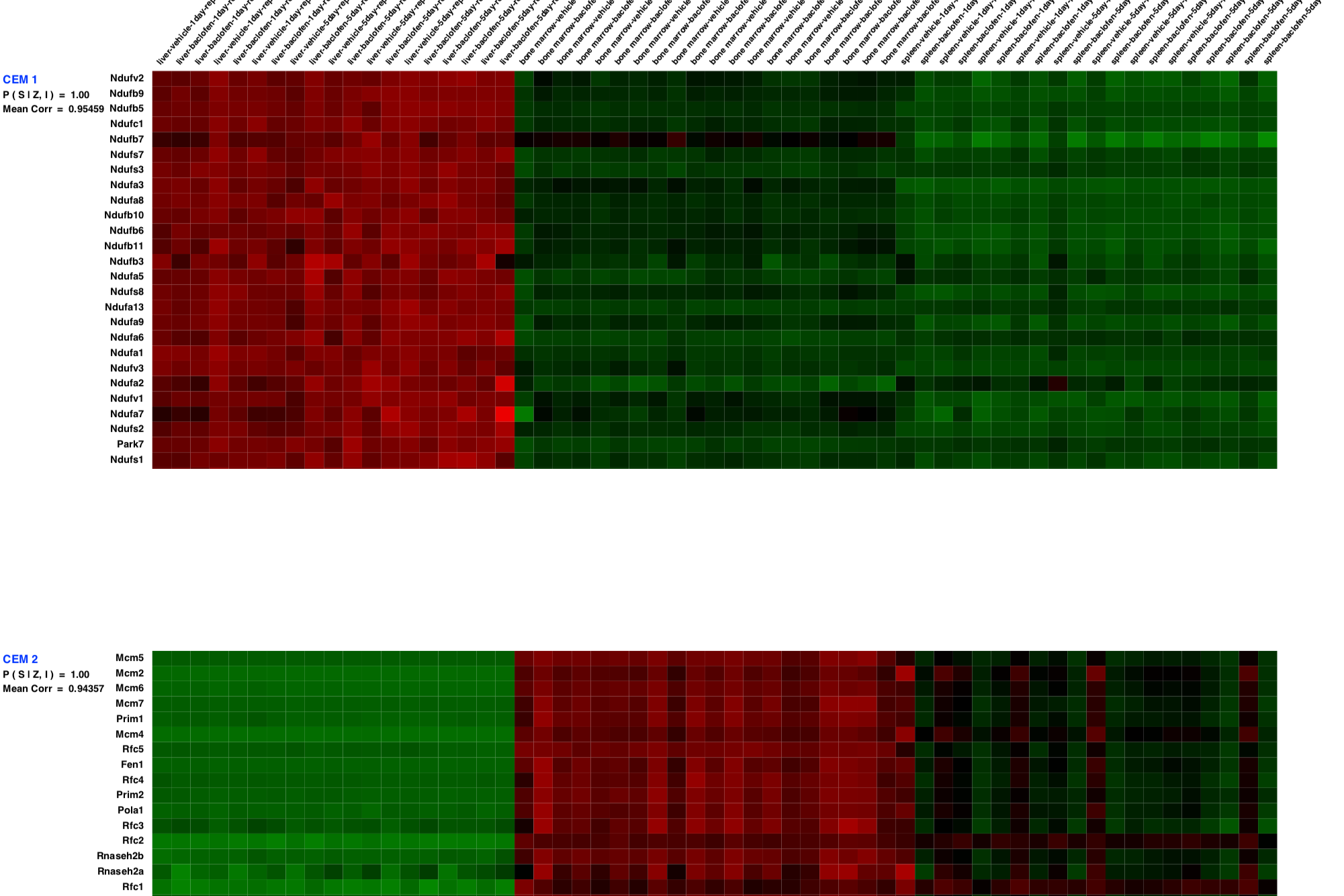} 
\caption{Expression levels of genes in CEM1 and CEM2 over samples in one dataset that is selected by both CEMs using CLIC. The top 26 rows correspond to genes in CEM1, and the bottom 16 rows correspond to genes in CEM2. Columns correspond to 49 experimental samples in the dataset. The expression level is in terms of Z scores. It ranges from -3 (green) to 3 (red).}
\label{fig: CEM_Partial}
\end{figure}
\clearpage
\end{landscape}
}

\section{Discussion} \label{sec: conclusion}
Modern high-throughput biomedical technologies have generated massive amount of data. The huge volume of data, complicated by the heterogeneity of data and the complexity of biological systems, posts increasing challenges for discovering hidden patterns and understanding biological functions. Bi-clustering is one way to deal with such challenges by identifying and clustering objects on a subset of features.
The goal of bi-clustering, when compared to clustering, is not only to cluster objects, but also to pinpoint important features and model them effectively for each cluster, which
reduces the impact of noise and helps unwind hidden structures in the data.

We have outlined a general Bayesian framework for tackling bi-clustering problems and described a few variants that target different types of data and attempt for different goals. For categorical data, we start with the basic model that clusters objects on a subset of features, which are assumed to have cluster-specific distributions in every cluster. The method can detect rare signals and has accurate clustering performance. This approach works well when data are  homogeneous, such as cell sub-types within one tissue/organ. We then extend the model to allow for different features to be used by different clusters. In this way, we are able to identify features that are differently distributed in a subset of clusters, but each follows a common feature-specific distribution in the remaining clusters. We then apply this general approach in a hierarchical fashion to identify bi-clusters in a step-wise manner. 
The sequence of node-splitting and the hierarchical structure show the relative heterogeneity among data and enable us to discover cluster-level relationships. As a useful and nontrivial twist of the standard bi-clustering framework, we lastly introduce a data-integration method that imposes a Bayesian bi-clustering structure on the correlation matrices derived from multiple datasets. We show that this approach can effectively extract important gene-module information by combining many publicly available gene expression datasets.

The true number of clusters in a dataset is often unknown. In our general framework, we have discussed the inference of $K$ based on its marginal posterior distribution. Another way of simultaneously inferring bi-cluster memberships and the total number of clusters is to employ a Dirichlet process mixture model. As $K$ changes throughout MCMC iterations, algorithms such as reversible jump MCMC and split-merge MCMC can be used. Readers can refer to methods in \cite{hoff2005subset, hoff2006model, varSelectionDP} for a detailed study on bi-clustering/clustering using Dirichlet process mixture models. 

The general framework can be easily extended to deal with data of mixed types, which often occur  when one tries to combine data from different sources measuring different aspects of a common set of objects. The assumption of independence among features in the general framework makes the analysis straightforward. However, in many cases, feature dependency cannot and should not be ignored. For example, genes in a pathway co-express in many biological functions. We have discussed one way to identify bi-clusters based on gene expression correlations. Another way to extend the general framework is to directly impose dependency structures on the observed data. For example, \cite{raftery2006variable} discussed the design and parameterization of the covariance matrix in Gaussian model-based clustering. A potential issue with this more idealistic approach is that the model can be too complex to allow for meaningful inferential results. More research along this direction is certainly warranted.

We have implemented the three Bayesian bi-clustering algorithms in R and C++, and it is available at https://github.com/HanY-H/BayesianBiClustering.
\newpage
\begin{appendix}

\section{Human genetic data description and BBC2's performance on the same data with shuffled columns}\label{app: HapMap Shuffled}
Table \ref{tab: Pop Description} summarizes the 11 populations in the human genetic data set analyzed in Sections \ref{sec: bbc2 hapmap} and \ref{sec: hbbc hapmap}.

\begin{table}[h!]
\begin{center}
\begin{tabular}{|c | l|c|}
\hline
Population & Population & Number of \\
Label & Description & Individuals\\
\hline
\hline
ASW & African ancestry in Southwest USA & 53\\
\hline
LWK & Luhya in Webuye, Kenya & 110\\
\hline
MKK & Maasai in Kinyawa, Kenya & 156\\
\hline
YRI & Yoruba in Ibadan, Nigeria & 147\\
\hline
GIH & Gujarati Indians in Houston, Texas & 101\\
\hline
MEX & Mexican ancestry in LA, California & 58\\
\hline
CHB & Han Chinese in Beijing, China & 137\\
\hline
CHD & Chinese in Metropolitan Denver, Colorado & 109\\
\hline
JPT & Japanese in Tokyo, Japan & 113\\
\hline
CEU & Utah residents with Northern and  & 112\\
& Western European ancestry& \\
\hline
TSI & Toscani in Italia & 102\\
\hline
\end{tabular}
\end{center}
\caption{A summary of the human genetic data used in Section \ref{sec: bbc2 hapmap} and \ref{sec: hbbc hapmap}.}
\label{tab: Pop Description}
\end{table}

In the following we present some results for the analysis of the shuffled data using BBC2. At each percentage of features (columns) to be shuffled (permuted), we repeat the shuffling procedure 10 times, i.e., for each level, 10 datasets are generated. Each dataset contains a randomly selected subset of  specified percentage of features being randomly permuted.
Table \ref{tab: Model2_summary_shuffled} shows the optimal number of clusters and clustering error rates of BBC2 at each pre-specified shuffling level for one shuffled dataset. Results for the rest 9 shuffled datasets at each level are very similar to Table \ref{tab: Model2_summary_shuffled}. Clustering results on the shuffled data at each pre-specified shuffling level are shown in Tables \ref{tab: Shuffle10} to \ref{tab: Shuffle90}.

\begin{table}[H]
\begin{center}
\begin{tabular}{|c|c | c|}
\hline
Percentage of shuffled columns & Number of Clusters & CE rate\\
\hline
10\% & 6 & 41.40\% \\
\hline
15\% & 6 & 41.57\% \\
\hline
20\% & 6 & 41.24\% \\
\hline
25\% & 6 & 41.32\% \\
\hline
30\% & 6 & 41.57\% \\
\hline
35\% & 6 & 41.57\% \\
\hline
40\% & 6 & 41.74\% \\
\hline
45\% & 6 & 41.40\% \\
\hline
50\% & 6 & 41.74\% \\
\hline
55\% & 6 & 41.74\% \\
\hline
60\% & 6 & 41.65\% \\
\hline
65\% & 6 & 41.65\% \\
\hline
70\% & 6 & 41.57\% \\
\hline
75\% & 5 & 53.51\% \\
\hline
80\% & 7 & 54.09\% \\
\hline
85\% & 4 & 57.76\% \\
\hline
90\% & 6 & 53.84\% \\
\hline
\end{tabular}    
\end{center}
\caption{A summary of BBC2's performance on the shuffled human genetic data at different shuffling levels.}
\label{tab: Model2_summary_shuffled}
\end{table}

\begin{table}[H]
\begin{center}
\begin{tabular}{|c |c c c c c c|}
\hline
Population & Cluster 1 & Cluster 2 & Cluster 3 & Cluster 4 & Cluster 5 & Cluster 6 \\
Abbreviation & & & & & & \\
\hline
\hline
ASW & 32& 21 & 0& 0& 0& 0\\
\hline
LWK & 110& 0& 0& 0& 0& 0\\
\hline
MKK & 3& 153& 0& 0& 0& 0\\
\hline
YRI & 147& 0& 0& 0& 0& 0\\
\hline
GIH & 0& 0& 101& 0& 0& 0\\
\hline
MEX & 0& 0& 0& 52& 6& 0\\
\hline
CHB & 0& 0& 0& 0& 0& 137\\
\hline
CHD & 0& 0& 0& 0& 0& 109\\
\hline
JPT &0& 0& 0& 0& 0& 113\\
\hline
CEU &0 &0 &0 &0 &112 & 0\\
\hline
TSI &0 &0 &0 & 0&102 & 0\\
\hline
\end{tabular}
\end{center}
\caption{BBC2’s clustering result on the shuffled data when 10\% of the columns are shuffled.}
\label{tab: Shuffle10}
\end{table}

\begin{table}[H]
\begin{center}
\begin{tabular}{|c |c c c c c c|}
\hline
Population & Cluster 1 & Cluster 2 & Cluster 3 & Cluster 4 & Cluster 5 & Cluster 6 \\
Abbreviation & & & & & & \\
\hline
\hline
ASW & 35& 18 & 0& 0& 0& 0\\
\hline
LWK & 110& 0& 0& 0& 0& 0\\
\hline
MKK & 5& 151& 0& 0& 0& 0\\
\hline
YRI & 147& 0& 0& 0& 0& 0\\
\hline
GIH & 0& 0& 101& 0& 0& 0\\
\hline
MEX & 0& 0& 0& 52& 6& 0\\
\hline
CHB & 0& 0& 0& 0& 0& 137\\
\hline
CHD & 0& 0& 0& 0& 0& 109\\
\hline
JPT &0& 0& 0& 0& 0& 113\\
\hline
CEU &0 &0 &0 &0 &112 & 0\\
\hline
TSI &0 &0 &0 & 0&102 & 0\\
\hline
\end{tabular}
\end{center}
\caption{BBC2’s clustering result on the shuffled data when 15\% of the columns are shuffled.}
\label{tab: Shuffle15}
\end{table}

\begin{table}[H]
\begin{center}
\begin{tabular}{|c |c c c c c c|}
\hline
Population & Cluster 1 & Cluster 2 & Cluster 3 & Cluster 4 & Cluster 5 & Cluster 6 \\
Abbreviation & & & & & & \\
\hline
\hline
ASW & 32& 21 & 0& 0& 0& 0\\
\hline
LWK & 110& 0& 0& 0& 0& 0\\
\hline
MKK & 2& 154& 0& 0& 0& 0\\
\hline
YRI & 147& 0& 0& 0& 0& 0\\
\hline
GIH & 0& 0& 101& 0& 0& 0\\
\hline
MEX & 0& 0& 0& 53& 5& 0\\
\hline
CHB & 0& 0& 0& 0& 0& 137\\
\hline
CHD & 0& 0& 0& 0& 0& 109\\
\hline
JPT &0& 0& 0& 0& 0& 113\\
\hline
CEU &0 &0 &0 &0 &112 & 0\\
\hline
TSI &0 &0 &0 & 0&102 & 0\\
\hline
\end{tabular}
\end{center}
\caption{BBC2’s clustering result on the shuffled data when 20\% of the columns are shuffled.}
\label{tab: Shuffle20}
\end{table}

\begin{table}[H]
\begin{center}
\begin{tabular}{|c |c c c c c c|}
\hline
Population & Cluster 1 & Cluster 2 & Cluster 3 & Cluster 4 & Cluster 5 & Cluster 6 \\
Abbreviation & & & & & & \\
\hline
\hline
ASW & 31& 22 & 0& 0& 0& 0\\
\hline
LWK & 110& 0& 0& 0& 0& 0\\
\hline
MKK & 4& 152& 0& 0& 0& 0\\
\hline
YRI & 147& 0& 0& 0& 0& 0\\
\hline
GIH & 0& 0& 101& 0& 0& 0\\
\hline
MEX & 0& 0& 0& 54& 4& 0\\
\hline
CHB & 0& 0& 0& 0& 0& 137\\
\hline
CHD & 0& 0& 0& 0& 0& 109\\
\hline
JPT &0& 0& 0& 0& 0& 113\\
\hline
CEU &0 &0 &0 &0 &112 & 0\\
\hline
TSI &0 &0 &0 & 0&102 & 0\\
\hline
\end{tabular}
\end{center}
\caption{BBC2’s clustering result on the shuffled data when 25\% of the columns are shuffled.}
\label{tab: Shuffle25}
\end{table}

\begin{table}[H]
\begin{center}
\begin{tabular}{|c |c c c c c c|}
\hline
Population & Cluster 1 & Cluster 2 & Cluster 3 & Cluster 4 & Cluster 5 & Cluster 6 \\
Abbreviation & & & & & & \\
\hline
\hline
ASW & 33& 20 & 0& 0& 0& 0\\
\hline
LWK & 110& 0& 0& 0& 0& 0\\
\hline
MKK & 6& 150& 0& 0& 0& 0\\
\hline
YRI & 147& 0& 0& 0& 0& 0\\
\hline
GIH & 0& 0& 101& 0& 0& 0\\
\hline
MEX & 0& 0& 0& 53& 3& 0\\
\hline
CHB & 0& 0& 0& 0& 0& 137\\
\hline
CHD & 0& 0& 0& 0& 0& 109\\
\hline
JPT &0& 0& 0& 0& 0& 113\\
\hline
CEU &0 &0 &0 &0 &112 & 0\\
\hline
TSI &0 &0 &0 & 0&102 & 0\\
\hline
\end{tabular}
\end{center}
\caption{BBC2’s clustering result on the shuffled data when 30\% of the columns are shuffled.}
\label{tab: Shuffle30}
\end{table}

\begin{table}[H]
\begin{center}
\begin{tabular}{|c |c c c c c c|}
\hline
Population & Cluster 1 & Cluster 2 & Cluster 3 & Cluster 4 & Cluster 5 & Cluster 6 \\
Abbreviation & & & & & & \\
\hline
\hline
ASW & 32& 21 & 0& 0& 0& 0\\
\hline
LWK & 110& 0& 0& 0& 0& 0\\
\hline
MKK & 4& 152& 0& 0& 0& 0\\
\hline
YRI & 147& 0& 0& 0& 0& 0\\
\hline
GIH & 0& 0& 101& 0& 0& 0\\
\hline
MEX & 0& 0& 0& 51& 7& 0\\
\hline
CHB & 0& 0& 0& 0& 0& 137\\
\hline
CHD & 0& 0& 0& 0& 0& 109\\
\hline
JPT &0& 0& 0& 0& 0& 113\\
\hline
CEU &0 &0 &0 &0 &112 & 0\\
\hline
TSI &0 &0 &0 & 0&102 & 0\\
\hline
\end{tabular}
\end{center}
\caption{BBC2’s clustering result on the shuffled data when 35\% of the columns are shuffled.}
\label{tab: Shuffle35}
\end{table}

\begin{table}[H]
\begin{center}
\begin{tabular}{|c |c c c c c c|}
\hline
Population & Cluster 1 & Cluster 2 & Cluster 3 & Cluster 4 & Cluster 5 & Cluster 6 \\
Abbreviation & & & & & & \\
\hline
\hline
ASW & 33& 20 & 0& 0& 0& 0\\
\hline
LWK & 110& 0& 0& 0& 0& 0\\
\hline
MKK & 5& 151& 0& 0& 0& 0\\
\hline
YRI & 147& 0& 0& 0& 0& 0\\
\hline
GIH & 0& 0& 101& 0& 0& 0\\
\hline
MEX & 0& 0& 0& 50& 8& 0\\
\hline
CHB & 0& 0& 0& 0& 0& 137\\
\hline
CHD & 0& 0& 0& 0& 0& 109\\
\hline
JPT &0& 0& 0& 0& 0& 113\\
\hline
CEU &0 &0 &0 &0 &112 & 0\\
\hline
TSI &0 &0 &0 & 0&102 & 0\\
\hline
\end{tabular}
\end{center}
\caption{BBC2’s clustering result on the shuffled data when 40\% of the columns are shuffled.}
\label{tab: Shuffle40}
\end{table}

\begin{table}[H]
\begin{center}
\begin{tabular}{|c |c c c c c c|}
\hline
Population & Cluster 1 & Cluster 2 & Cluster 3 & Cluster 4 & Cluster 5 & Cluster 6 \\
Abbreviation & & & & & & \\
\hline
\hline
ASW & 31& 22 & 0& 0& 0& 0\\
\hline
LWK & 110& 0& 0& 0& 0& 0\\
\hline
MKK & 3& 153& 0& 0& 0& 0\\
\hline
YRI & 147& 0& 0& 0& 0& 0\\
\hline
GIH & 0& 0& 101& 0& 0& 0\\
\hline
MEX & 0& 0& 0& 52& 6& 0\\
\hline
CHB & 0& 0& 0& 0& 0& 137\\
\hline
CHD & 0& 0& 0& 0& 0& 109\\
\hline
JPT &0& 0& 0& 0& 0& 113\\
\hline
CEU &0 &0 &0 &0 &112 & 0\\
\hline
TSI &0 &0 &0 & 0&102 & 0\\
\hline
\end{tabular}
\end{center}
\caption{BBC2’s clustering result on the shuffled data when 45\% of the columns are shuffled.}
\label{tab: Shuffle45}
\end{table}

\begin{table}[H]
\begin{center}
\begin{tabular}{|c |c c c c c c|}
\hline
Population & Cluster 1 & Cluster 2 & Cluster 3 & Cluster 4 & Cluster 5 & Cluster 6 \\
Abbreviation & & & & & & \\
\hline
\hline
ASW & 33& 20 & 0& 0& 0& 0\\
\hline
LWK & 110& 0& 0& 0& 0& 0\\
\hline
MKK & 9& 147& 0& 0& 0& 0\\
\hline
YRI & 147& 0& 0& 0& 0& 0\\
\hline
GIH & 0& 0& 101& 0& 0& 0\\
\hline
MEX & 0& 0& 0& 54& 4& 0\\
\hline
CHB & 0& 0& 0& 0& 0& 137\\
\hline
CHD & 0& 0& 0& 0& 0& 109\\
\hline
JPT &0& 0& 0& 0& 0& 113\\
\hline
CEU &0 &0 &0 &0 &112 & 0\\
\hline
TSI &0 &0 &0 & 0&102 & 0\\
\hline
\end{tabular}
\end{center}
\caption{BBC2’s clustering result on the shuffled data when 50\% of the columns are shuffled.}
\label{tab: Shuffle50}
\end{table}

\begin{table}[H]
\begin{center}
\begin{tabular}{|c |c c c c c c|}
\hline
Population & Cluster 1 & Cluster 2 & Cluster 3 & Cluster 4 & Cluster 5 & Cluster 6 \\
Abbreviation & & & & & & \\
\hline
\hline
ASW & 32& 21 & 0& 0& 0& 0\\
\hline
LWK & 110& 0& 0& 0& 0& 0\\
\hline
MKK & 7& 149& 0& 0& 0& 0\\
\hline
YRI & 147& 0& 0& 0& 0& 0\\
\hline
GIH & 0& 0& 101& 0& 0& 0\\
\hline
MEX & 0& 0& 0& 52& 6& 0\\
\hline
CHB & 0& 0& 0& 0& 0& 137\\
\hline
CHD & 0& 0& 0& 0& 0& 109\\
\hline
JPT &0& 0& 0& 0& 0& 113\\
\hline
CEU &0 &0 &0 &0 &112 & 0\\
\hline
TSI &0 &0 &0 & 0&102 & 0\\
\hline
\end{tabular}
\end{center}
\caption{BBC2’s clustering result on the shuffled data when 55\% of the columns are shuffled.}
\label{tab: Shuffle55}
\end{table}

\begin{table}[H]
\begin{center}
\begin{tabular}{|c |c c c c c c|}
\hline
Population & Cluster 1 & Cluster 2 & Cluster 3 & Cluster 4 & Cluster 5 & Cluster 6 \\
Abbreviation & & & & & & \\
\hline
\hline
ASW & 36& 17 & 0& 0& 0& 0\\
\hline
LWK & 110& 0& 0& 0& 0& 0\\
\hline
MKK & 8& 148& 0& 0& 0& 0\\
\hline
YRI & 147& 0& 0& 0& 0& 0\\
\hline
GIH & 0& 0& 101& 0& 0& 0\\
\hline
MEX & 0& 0& 0& 54& 4& 0\\
\hline
CHB & 0& 0& 0& 0& 0& 137\\
\hline
CHD & 0& 0& 0& 0& 0& 109\\
\hline
JPT &0& 0& 0& 0& 0& 113\\
\hline
CEU &0 &0 &0 &0 &112 & 0\\
\hline
TSI &0 &0 &0 & 0&102 & 0\\
\hline
\end{tabular}
\end{center}
\caption{BBC2’s clustering result on the shuffled data when 60\% of the columns are shuffled.}
\label{tab: Shuffle60}
\end{table}

\begin{table}[H]
\begin{center}
\begin{tabular}{|c |c c c c c c|}
\hline
Population & Cluster 1 & Cluster 2 & Cluster 3 & Cluster 4 & Cluster 5 & Cluster 6 \\
Abbreviation & & & & & & \\
\hline
\hline
ASW & 32& 21 & 0& 0& 0& 0\\
\hline
LWK & 110& 0& 0& 0& 0& 0\\
\hline
MKK & 4& 152& 0& 0& 0& 0\\
\hline
YRI & 147& 0& 0& 0& 0& 0\\
\hline
GIH & 0& 0& 101& 0& 0& 0\\
\hline
MEX & 0& 0& 0& 50& 8& 0\\
\hline
CHB & 0& 0& 0& 0& 0& 137\\
\hline
CHD & 0& 0& 0& 0& 0& 109\\
\hline
JPT &0& 0& 0& 0& 0& 113\\
\hline
CEU &0 &0 &0 &0 &112 & 0\\
\hline
TSI &0 &0 &0 & 0&102 & 0\\
\hline
\end{tabular}
\end{center}
\caption{BBC2’s clustering result on the shuffled data when 65\% of the columns are shuffled.}
\label{tab: Shuffle65}
\end{table}

\begin{table}[H]
\begin{center}
\begin{tabular}{|c |c c c c c c|}
\hline
Population & Cluster 1 & Cluster 2 & Cluster 3 & Cluster 4 & Cluster 5 & Cluster 6 \\
Abbreviation & & & & & & \\
\hline
\hline
ASW & 36& 17 & 0& 0& 0& 0\\
\hline
LWK & 110& 0& 0& 0& 0& 0\\
\hline
MKK & 5& 151& 0& 0& 0& 0\\
\hline
YRI & 147& 0& 0& 0& 0& 0\\
\hline
GIH & 0& 0& 101& 0& 0& 0\\
\hline
MEX & 0& 0& 0& 52& 6& 0\\
\hline
CHB & 0& 0& 0& 0& 0& 137\\
\hline
CHD & 0& 0& 0& 0& 0& 109\\
\hline
JPT &0& 0& 0& 0& 0& 113\\
\hline
CEU &0 &0 &0 &0 &112 & 0\\
\hline
TSI &0 &0 &0 & 0&102 & 0\\
\hline
\end{tabular}
\end{center}
\caption{BBC2’s clustering result on the shuffled data when 70\% of the columns are shuffled.}
\label{tab: Shuffle70}
\end{table}

\begin{table}[H]
\begin{center}
\begin{tabular}{|c |c c c c c|}
\hline
Population & Cluster 1 & Cluster 2 & Cluster 3 & Cluster 4 & Cluster 5 \\
Abbreviation & & & & & \\
\hline
\hline
ASW & 53& 0 & 0& 0& 0\\
\hline
LWK & 110& 0& 0& 0& 0\\
\hline
MKK & 156& 0& 0& 0& 0\\
\hline
YRI & 147& 0& 0& 0& 0\\
\hline
GIH & 0& 100& 1& 0& 0\\
\hline
MEX & 0& 0& 6& 52& 0\\
\hline
CHB & 0& 0& 0& 0& 137\\
\hline
CHD & 0& 0& 0& 0& 109\\
\hline
JPT &0& 0& 0& 0& 113\\
\hline
CEU &0 &0 &112 & 0 &0\\
\hline
TSI &0 &0 &102 & 0 &0\\
\hline
\end{tabular}
\end{center}
\caption{BBC2’s clustering result on the shuffled data when 75\% of the columns are shuffled.}
\label{tab: Shuffle75}
\end{table}

\begin{table}[H]
\begin{center}
\begin{tabular}{|c |c c c c c c c|}
\hline
Population & Cluster 1 & Cluster 2 & Cluster 3 & Cluster 4 & Cluster 5 & Cluster 6 & Cluster 7\\
Abbreviation & & & & & & & \\
\hline
\hline
ASW & 53& 0 &0 & 0& 0& 0& 0\\
\hline
LWK & 110& 0& 0& 0& 0& 0 &0\\
\hline
MKK &156 &0 &0 & 0& 0& 0& 0\\
\hline
YRI & 147 &0 & 0& 0& 0& 0& 0\\
\hline
GIH & 0& 101& 0& 0& 0 &0 &0 \\
\hline
MEX & 0& 0& 44& 10 &2 & 2 &0\\
\hline
CHB & 0& 0& 0& 0& 0& 0& 137\\
\hline
CHD & 0& 0& 0& 0& 0& 0&  109\\
\hline
JPT &0& 0& 0& 0& 0& 0& 113\\
\hline
CEU &0 &0 &0 &112 & 0 &0 &0\\
\hline
TSI &0 &0 &0 &102 & 0 &0 &0\\
\hline
\end{tabular}
\end{center}
\caption{BBC2’s clustering result on the shuffled data when 80\% of the columns are shuffled.}
\label{tab: Shuffle80}
\end{table}

\begin{table}[H]
\begin{center}
\begin{tabular}{|c |c c c c|}
\hline
Population & Cluster 1 & Cluster 2 & Cluster 3 & Cluster 4 \\
Abbreviation & & & & \\
\hline
\hline
ASW & 53& 0 & 0& 0\\
\hline
LWK & 110& 0& 0& 0\\
\hline
MKK & 156& 0& 0& 0\\
\hline
YRI & 147& 0& 0& 0\\
\hline
GIH & 0& 101& 0& 0\\
\hline
MEX & 0& 50& 8& 0\\
\hline
CHB & 0& 0& 0& 137\\
\hline
CHD & 0& 0& 0& 109\\
\hline
JPT &0& 0& 0& 113\\
\hline
CEU &0 &0 &112 & 0\\
\hline
TSI &0 &0 &102  &0\\
\hline
\end{tabular}
\end{center}
\caption{BBC2’s clustering result on the shuffled data when 85\% of the columns are shuffled.}
\label{tab: Shuffle85}
\end{table}

\begin{table}[H]
\begin{center}
\begin{tabular}{|c |c c c c c c|}
\hline
Population & Cluster 1 & Cluster 2 & Cluster 3 & Cluster 4 & Cluster 5 & Cluster 6 \\
Abbreviation & & & & & & \\
\hline
\hline
ASW & 53& 0 & 0& 0& 0& 0\\
\hline
LWK & 110& 0& 0& 0& 0& 0\\
\hline
MKK & 156& 0& 0& 0& 0& 0\\
\hline
YRI & 147& 0& 0& 0& 0& 0\\
\hline
GIH & 0& 101& 0& 0& 0& 0\\
\hline
MEX & 0& 0& 47& 5& 6& 0\\
\hline
CHB & 0& 0& 0& 0& 0& 137\\
\hline
CHD & 0& 0& 0& 0& 0& 109\\
\hline
JPT &0& 0& 0& 0& 0& 113\\
\hline
CEU &0 &0 &0 &0 &112 & 0\\
\hline
TSI &0 &0 &0 & 0&102 & 0\\
\hline
\end{tabular}
\end{center}
\caption{BBC2’s clustering result on the shuffled data when 90\% of the columns are shuffled.}
\label{tab: Shuffle90}
\end{table}

Table \ref{tab: falsePos_shuffled} records false positive rates for shuffled features at different shuffling levels computed based on BBC2's outputs. Since for each randomly selected feature  that has its observations randomly shuffled, we expect the feature to be a pure noise column and not to be selected by any clusters. Low false positive rates indicate that BBC2 is powerful to detect such noises and not use them for clustering.

\begin{table}[H]
\begin{center}
\begin{tabular}{|c c|c c|c c|}
\hline
Shuffling & False Positive&Shuffling  & False Positive &Shuffling & False Positive\\
Level & Rate & Level & Rate & Level & Rate\\
\hline\hline
10\% & 3.55\% & 40\% & 3.79\% & 70\% & 3.15\%\\
\hline
15\% & 4.42\% &45\% & 3.53\% & 75\% & 6.35\% \\
\hline
20\% & 3.32\% & 50\% & 3.22\% & 80\% & 10.23\%\\
\hline
25\% & 3.22\% & 55\% & 6.21\% & 85\% & 9.93\%\\
\hline
30\% & 3.63\% & 60\% & 3.16\% & 90\% & 7.22\% \\
\hline
35\% & 3.39\% & 65\% & 6.02\% & & \\
\hline
\end{tabular}
\end{center}
\caption{False positive rates among shuffled columns at different shuffling levels computed from BBC2's outputs.}
\label{tab: falsePos_shuffled}
\end{table}

\newpage
\section{BBC2's clustering results on the human genetic data with different thresholds on pairwise correlation}\label{app: Hapmap_r2}
By setting the upper bound for $r^2$ to be $10^{-4}$, 4,518 SNPs are included in the analysis. BBC2 still selects the optimal number of clusters to be 6, and its clustering result is shown in Table \ref{tab: HapMap_r4}. A summary of feature heterogeneity is shown in Table \ref{tab: SNP overlap_r4}.

\begin{table}[h!]
\begin{center}
\begin{tabular}{|c |c c c c c c|}
\hline
Population & Cluster 1 & Cluster 2 & Cluster 3 & Cluster 4 & Cluster 5 & Cluster 6 \\
Abbreviation & & & & & & \\
\hline
\hline
ASW & 32& 21 & 0& 0& 0& 0\\
\hline
LWK & 110& 0& 0& 0& 0& 0\\
\hline
MKK & 4& 152& 0& 0& 0& 0\\
\hline
YRI & 147& 0& 0& 0& 0& 0\\
\hline
GIH & 0& 0& 101& 0& 0& 0\\
\hline
MEX & 0& 0& 0& 53& 5& 0\\
\hline
CHB & 0& 0& 0& 0& 0& 137\\
\hline
CHD & 0& 0& 0& 0& 0& 109\\
\hline
JPT &0& 0& 0& 0& 0& 113\\
\hline
CEU &0 &0 &0 &0 &112 & 0\\
\hline
TSI &0 &0 &0 & 0&102 & 0\\
\hline
\end{tabular}
\end{center}
\caption{BBC2's clustering result when the upper bound for $r^2$ is $10^{-4}$.}
\label{tab: HapMap_r4}
\end{table}

\begin{table}[h!]
\begin{center}
\begin{tabular}{|l|c c c c c c|}
\hline
Number of Cluster-Specific Distributions &0 & 1 & 2 & 3 &4 &6 \\
\hline
Number of SNPs & 179 & 689 & 1553 & 1608 & 480 & 9 \\
\hline
\end{tabular}
\end{center}
\caption{BBC2's estimate on numbers of SNPs that have the specified numbers of cluster-specific distributions for the human genetic dataset with $r^2=10^{-4}$. }
\label{tab: SNP overlap_r4}
\end{table}

By setting the upper bound for $r^2$ to be $10^{-2}$, 14,840 SNPs are included in the anlaysis, which is more than 3 times the number of SNPs included when the upper bound is $10^{-4}$. These additional SNPs seem to help in better clustering ASW samples, with most of them being in  cluster 1 and only 5 individuals in cluster 2. However, it also introduces additional noises, resulting in more MEX individuals being separated from its main cluster, i.e., cluster 4, and being mixed with the European cluster. The clustering and feature selection results are shown in Tables \ref{tab: HapMap_r2} and \ref{tab: SNP overlap_r2}.

\begin{table}[h!]
\begin{center}
\begin{tabular}{|c |c c c c c c|}
\hline
Population & Cluster 1 & Cluster 2 & Cluster 3 & Cluster 4 & Cluster 5 & Cluster 6 \\
Abbreviation & & & & & & \\
\hline
\hline
ASW & 48& 5 & 0& 0& 0& 0\\
\hline
LWK & 110& 0& 0& 0& 0& 0\\
\hline
MKK & 3& 153& 0& 0& 0& 0\\
\hline
YRI & 147& 0& 0& 0& 0& 0\\
\hline
GIH & 0& 0& 101& 0& 0& 0\\
\hline
MEX & 0& 0& 0& 46& 12& 0\\
\hline
CHB & 0& 0& 0& 0& 0& 137\\
\hline
CHD & 0& 0& 0& 0& 0& 109\\
\hline
JPT &0& 0& 0& 0& 0& 113\\
\hline
CEU &0 &0 &0 &0 &112 & 0\\
\hline
TSI &0 &0 &0 & 0&102 & 0\\
\hline
\end{tabular}
\end{center}
\caption{BBC2's clustering result when the upper bound for $r^2$ is $10^{-2}$.}
\label{tab: HapMap_r2}
\end{table}

\begin{table}[h!]
\begin{center}
\begin{tabular}{|l|c c c c c c|}
\hline
Number of Cluster-Specific Distributions &0 & 1 & 2 & 3 &4 &6\\
\hline
Number of SNPs & 4991 & 2554 & 3549 & 3029 & 706 & 11 \\
\hline
\end{tabular}
\end{center}
\caption{BBC2's estimate on numbers of SNPs that have the specified numbers of cluster-specific distributions for the human genetic dataset with $r^2=10^{-2}$. }
\label{tab: SNP overlap_r2}
\end{table}

\newpage
\section{HBBC's  results on the human genetic data with different minimum node sizes}\label{app: min node size}
\begin{figure}[H]
\centering
\includegraphics[width=4.5in,height=2.5in]{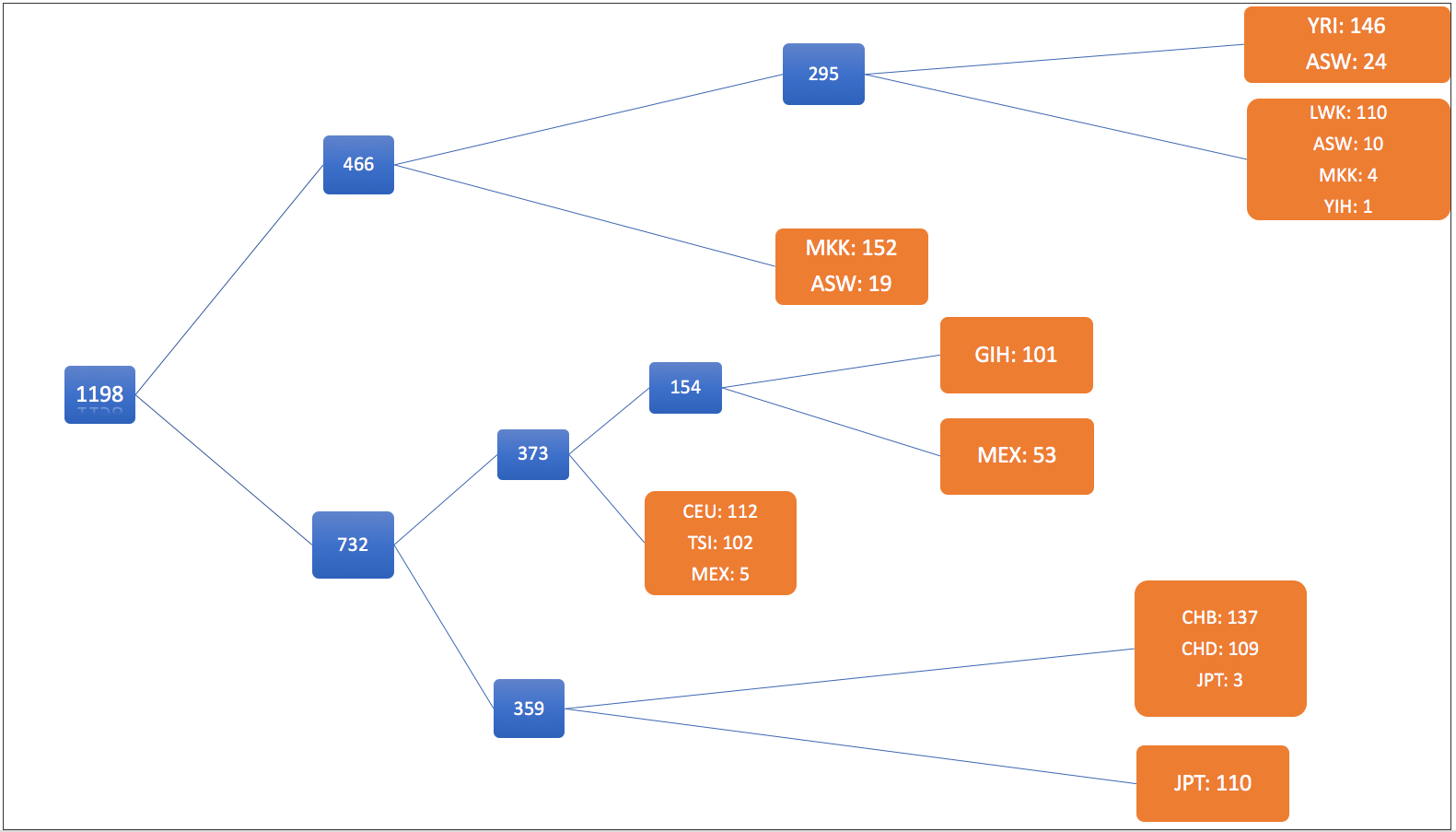}
\caption{HBBC's estimation result on the human genetic data, minimum node size=40. Tree and node interpretations follow from Section \ref{sec: hbbc hapmap}.}
\label{fig: Tree_min40}
\end{figure}

\begin{figure}[H]
\centering
\includegraphics[width=4.5in,height=2.5in]{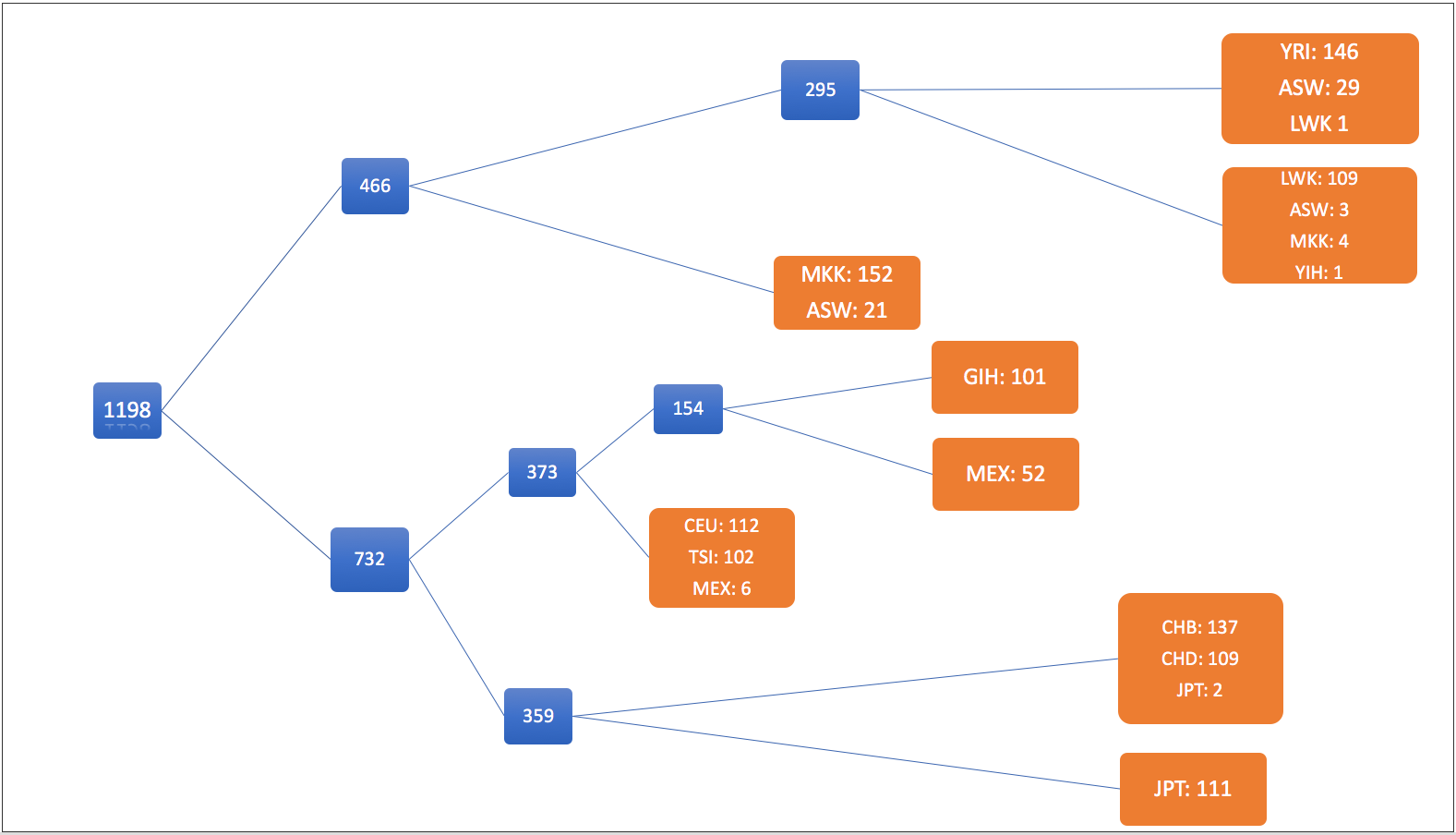}
\caption{HBBC's estimation result on the human genetic data, minimum node size=30. Tree and node interpretations follow from Section \ref{sec: hbbc hapmap}.}
\label{fig: Tree_min30}
\end{figure}

\begin{figure}[H]
\centering
\includegraphics[width=4.5in,height=2.5in]{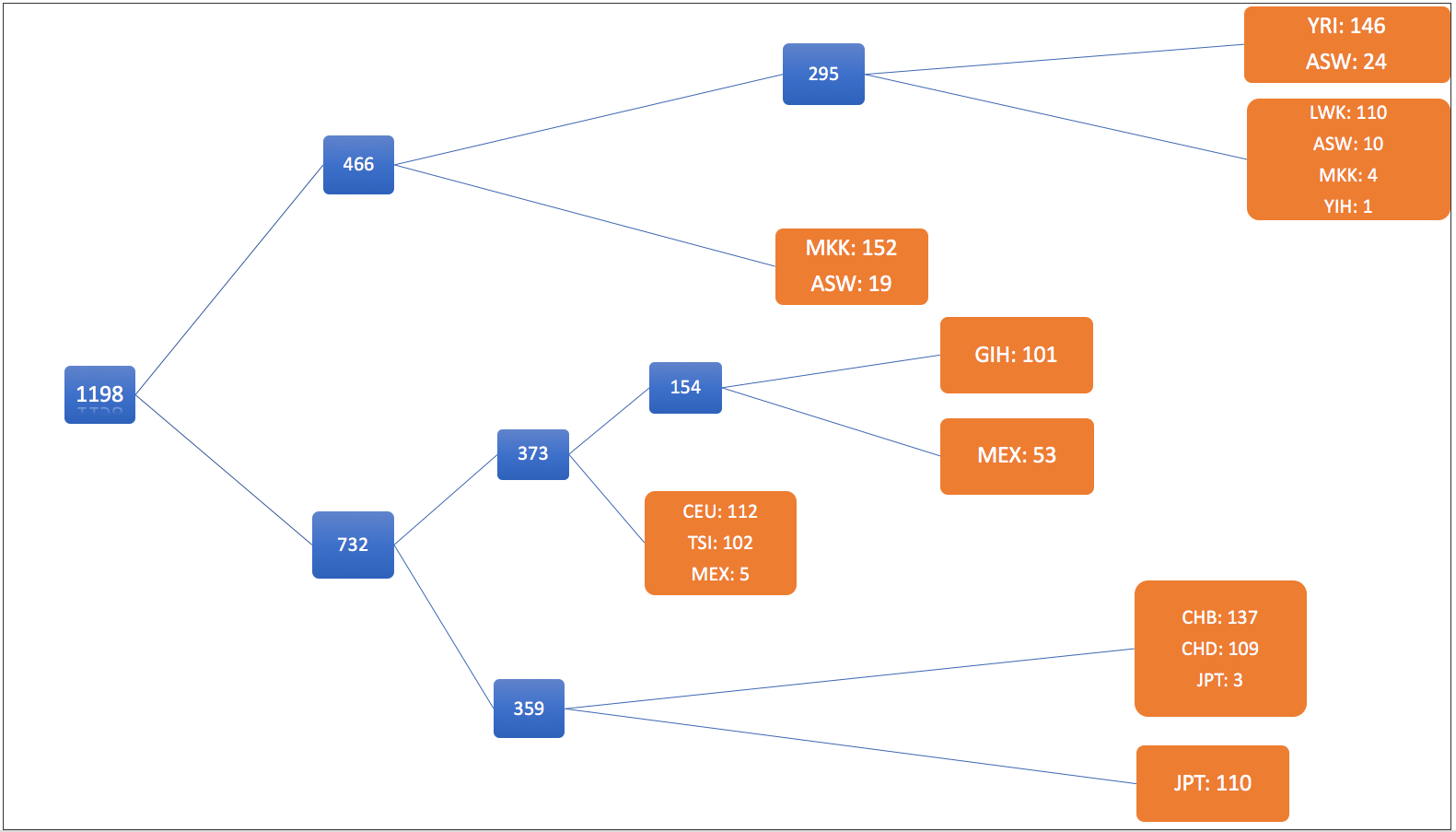}
\caption{HBBC's estimation result on the human genetic data, minimum node size=20. Tree and node interpretations follow from Section \ref{sec: hbbc hapmap}.}
\label{fig: Tree_min20}
\end{figure}

Figures \ref{fig: Tree_min40}-\ref{fig: Tree_min20} show HBBC's  results on the human genetic dataset analyzed in Section \ref{sec: hbbc hapmap} with the minimum node size decreasing from 40 to 5. HBBC is quite robust to the minimum node size specification.

\begin{figure}[H]
\centering
\includegraphics[width=4.5in,height=2.5in]{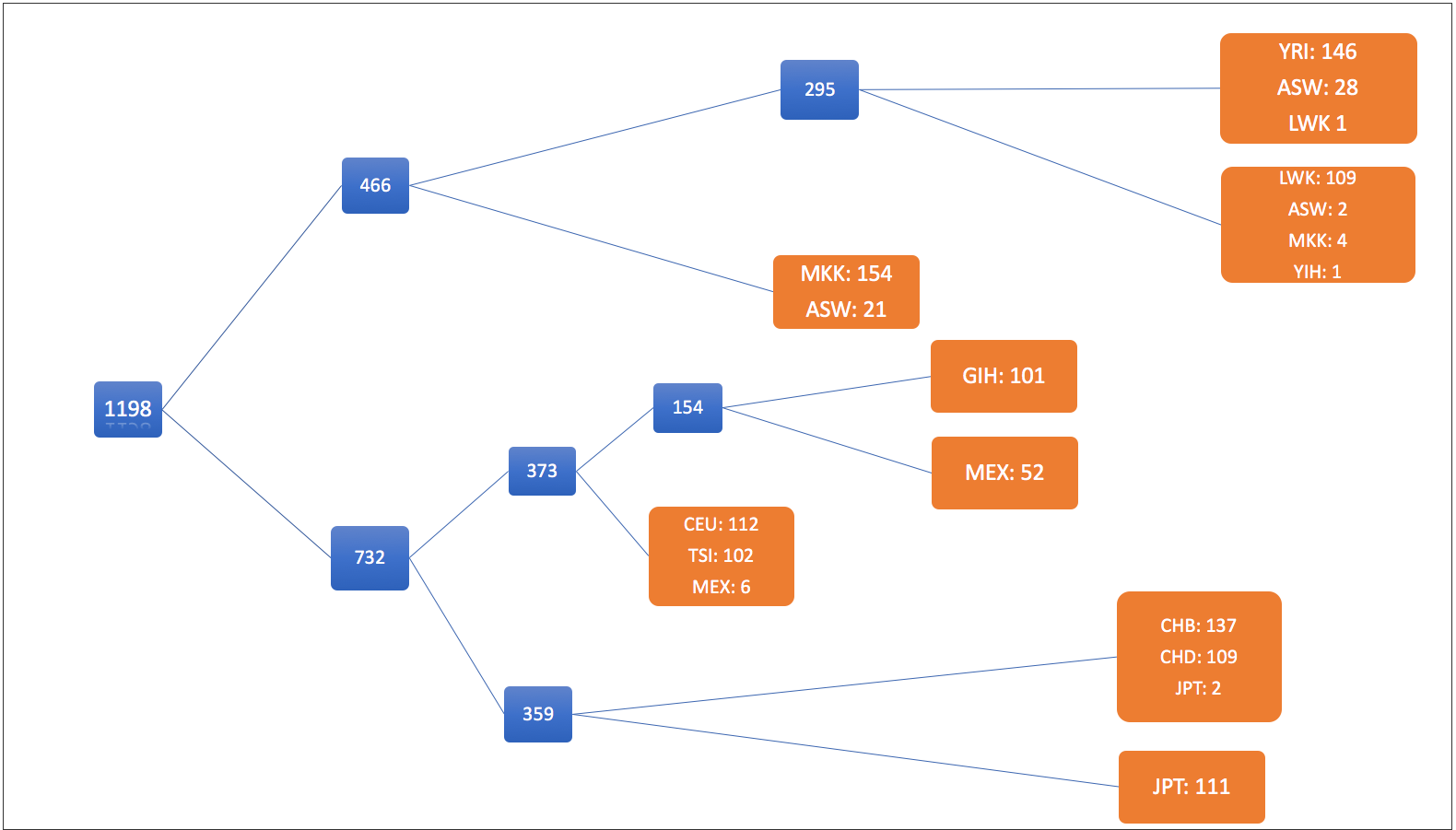}
\caption{HBBC's estimation result on the human genetic data, minimum node size=5. Tree and node interpretations follow from Section \ref{sec: hbbc hapmap}.}
\label{fig: Tree_min5}
\end{figure}

\section{Analysis of hair-related SNPs}\label{sec:hair}
We first identify a larger pool of SNPs from our dataset. This pool includes both the 4,217 SNPs in the final dataset fed to HBBC and SNPs that are close to and have high correlations with the 4,217 SNPs. 
The resulting pool contains 34,052 SNPs. In each step, we partition the 34,052 SNPs into two sets: a target set that includes SNPs that are either selected in this step or are highly correlated and near a selected SNP, and a non-target set that contains SNPs that are either not selected in this step or are highly correlated and near a not-selected SNP. Any overlap between the two sets are removed from both sets. We obtain a list of 93 hair and hair color related SNPs from Phenotype-Genotype Integrator website. 4 of such SNPs are in the pool of 34,052 SNPs. We calculate the odds ratio of the following two: being a traits-related SNP and in the target set over being a traits-related SNP but in the non-target set, and being in the target set over being in the non-target set. In step 1, i.e., separating the African populations from the rest of the populations, all 4 SNPs are in the target set. The odds ratios are 2.92, 1.83 and 1.29 in step 5, step 3 and step 2, respectively. All of these steps involve separating populations in different continents. While in step 4, separating the MKK population from the other African sub-populations, the odds ratio is 0.82, indicating hair and hair color become less distinguishable at this stage. The odds ratios are 0, i.e., none of the traits-related SNPs is selected in step 6 and 7: separating sub-populations in Africa and sub-populations in east Asia. This may suggest that specific aspects in hair and hair color related traits that are associated with the 4 SNPs are not distinguishable between these closely related sub-populations.

\section{More details on BBC for data integration}\label{app: clic}

We  conjugate a Normal-Inverse-Gamma prior for $\theta_{d,k}$ and $\sigma_{d,k}^2$ in Section~\ref{Integrative}
\begin{align*}
\theta_{d,k} & \sim N\left(\mu_{\theta},\sigma_{d,k}^{2}/\kappa_{\theta}\right),\quad k=1,\dots,K,\\
\sigma_{d,k}^{2} & \sim\text{Inv-Gamma}\left(\alpha_{\sigma},\beta_{\sigma}\right),\quad k=1,\dots,K,
\end{align*}
where $\mu_\theta$, $\kappa_\theta$, $\alpha_\sigma$ and $\beta_\sigma$ are hyper-parameters. 


To enable the Gibbs sampler, we integrate out $\bm{\theta}$ and $\bm{\sigma}$ to obtain the marginal likelihood function for $\bm{C}$ and $\bm{S}$,
\begin{equation*}
\begin{aligned} & P\left(\bY \mid\boldsymbol{C},\boldsymbol{S}, K \right)\\
= & \int\int P\left(\bY \mid\boldsymbol{\theta},\boldsymbol{\sigma},\boldsymbol{C},\boldsymbol{S}, K \right)P\left(\boldsymbol{\theta}\mid\boldsymbol{C}, K\right)P\left(\boldsymbol{\sigma}\mid\boldsymbol{C}, K \right)d\boldsymbol{\theta}d\boldsymbol{\sigma}\\
= & \prod_{d=1}^{p}\left[\prod_{k=1}^{K}\left\{ \frac{\left\{ \beta_{\sigma}^{\alpha_{\sigma}}\left(2\pi\right)^{-\frac{n_{k}}{2}}/\sqrt{1+\kappa_{\theta}^{-1}n_{k}}\right\} \cdot\left\{ \Gamma\left(\alpha_{\sigma}+\frac{n_{k}}{2}\right)/\Gamma\left(\alpha_{\sigma}\right)\right\} }{\left(\beta_{\sigma}+\frac{1}{2}\left(\underset{i<j:c_{i}=c_{j}=k}{\sum}Y_{d,i,j}^{2}+\kappa_{\theta}\mu_{\theta}^{2}-\frac{\left(\kappa_{\theta}\mu_{\theta}+\underset{i<j:c_{i}=c_{j}=k}{\sum}Y_{d,i,j}\right)^{2}}{\kappa_{\theta}+n_{k}}\right)\right)^{\alpha_{\sigma}+\frac{n_{k}}{2}}}\right\} ^{S_{d,k}}\right]\\
 & \cdot\prod_{d=1}^{p}\left[\prod_{i<j:\,c_{i}\neq c_{j}}\frac{\exp\left\{ -\frac{\left(Y_{d,i,j}-\theta_{d,0}\right)^{2}}{2\sigma_{d,0}^{2}}\right\} }{\sqrt{2\pi\sigma_{d,0}^{2}}}\right]^{S_{d,k}}\cdot\prod_{d=1}^{p}\left[\prod_{i<j}\frac{\exp\left\{ -\frac{\left(Y_{d,i,j}-\theta_{d,0}\right)^{2}}{2\sigma_{d,0}^{2}}\right\} }{\sqrt{2\pi\sigma_{d,0}^{2}}}\right]^{1-S_{d,k}},
\end{aligned}
\end{equation*}
where $n_k$ denotes the number of gene pairs in cluster $k$.
We can also integrate out $\bm{S}$ from the likelihood function $P\left(\bY \mid\boldsymbol{C},\boldsymbol{S}, K \right)$ to obtain the marginal distribution of $\bm{C}$, so that the Gibbs sampler only cycles through $C_1,\dots,C_n$:
\begin{equation*}
\begin{aligned} & \;\;P\left(\bY \mid\boldsymbol{C}, K \right)\\
= & \:\sum_{\boldsymbol{S}_{1}\in\left\{ 0,1\right\} ^{K}}P\left(\boldsymbol{S}_{1}\mid\boldsymbol{C}, K \right)\cdots\sum_{\boldsymbol{S}_{p}\in\left\{ 0,1\right\} ^{K}}P\left(\boldsymbol{S}_{p}\mid\boldsymbol{C}, K \right)\,P\left(\bY \mid\boldsymbol{C},\boldsymbol{S}, K\right)\\
= & \:\prod_{d=1}^{p}\left[\prod_{k=1}^{K}\left[\left(\pi_{S}\right)\left(\frac{\left\{ \beta_{\sigma}^{\alpha_{\sigma}}\left(2\pi\right)^{-\frac{n_{k}}{2}}/\sqrt{1+\kappa_{\theta}^{-1}n_{k}}\right\} \cdot\left\{ \Gamma\left(\alpha_{\sigma}+\frac{n_{k}}{2}\right)/\Gamma\left(\alpha_{\sigma}\right)\right\} }{\left(\beta_{\sigma}+\frac{1}{2}\left(\underset{i<j:c_{i}=c_{j}=k}{\sum}Y_{d,i,j}^{2}+\kappa_{\theta}\mu_{\theta}^{2}-\frac{\left(\kappa_{\theta}\mu_{\theta}+\underset{i<j:c_{i}=c_{j}=k}{\sum}Y_{d,i,j}\right)^{2}}{\kappa_{\theta}+n_{k}}\right)\right)^{\alpha_{\sigma}+\frac{n_{k}}{2}}}\right)\right.\right.\\
 & \left.\;\left.+\left(1-\pi_{S}\right)\left(\prod_{i<j:c_{i}=c_{j}=k}\frac{\exp\left\{ -\frac{\left(Y_{d,i,j}-\theta_{d,0}\right)^{2}}{2\sigma_{d,0}^{2}}\right\} }{\sqrt{2\pi\sigma_{d,0}^{2}}}\right)\right]\cdot\left(\prod_{i<j:c_{i}\neq c_{j}}\frac{\exp\left\{ -\frac{\left(Y_{d,i,j}-\theta_{d,0}\right)^{2}}{2\sigma_{d,0}^{2}}\right\} }{\sqrt{2\pi\sigma_{d,0}^{2}}}\right)\right].
\end{aligned}
\end{equation*}

We run the sampler for a range of cluster numbers, $K=1,\dots,\mathcal{K}$, and obtain the point estimator of number of clusters $\hat{K}$ and cluster assignment $\hat{\bm{C}}$ by the MAP estimator. Let $\bm{C}_{\left(K\right)}^{\left(1\right)},\dots,\bm{C}_{\left(K\right)}^{\left(M\right)}$ denote MCMC samples of $\bm{C}$ with the number of clusters $K$. The MAP estimator for $K$ and $\bm{C}$ is defined as
\begin{equation}\label{eq:KC}
\left(\hat{K},\hat{\bm{C}} \right)=\underset{K,\bm{C}_{\left(K\right)}^{\left(m\right)}:\:K=1,\dots,\mathcal{K};\,m=1,\dots,M}{\arg\max}\;\:P\left(\bm{C}_{\left(K\right)}^{\left(m\right)}\mid\bY \right).
\end{equation}

\end{appendix}

\section*{Acknowledgements}
This research was supported in part by NSF Grants  DMS-1903139 and DMS-2015411.

\bibliographystyle{chicago}
\bibliography{main}
\end{document}